\documentclass[abstract]{aa}
\usepackage[lofdepth,lotdepth]{subfig}
\usepackage{graphicx}
\usepackage{threeparttable}
\usepackage{capt-of}
\usepackage{txfonts}
\usepackage{natbib}
\usepackage{float}
\usepackage{hyperref}
\usepackage[switch]{lineno} 
\usepackage{lineno}
\usepackage{multirow}
\usepackage{xcolor}
\usepackage{textcomp}
\usepackage{orcidlink}

\newcommand{\msun}{{M$_\odot$}}
\newcommand{\mJypsf}{mJy PSF$^{-1}$}
\newcommand{\Jypsf}{Jy PSF$^{-1}$}

\newcommand{\uJyarcsecsq}{$\mu$Jy arcsec$^{-2}$}

\newcommand{\whz}{$\,$W$\,$Hz$^{-1}$}

\newcommand{\kmsmpc}{$\,$km$\,$s$^{-1}$Mpc$^{-1}$}

\bibpunct{(}{)}{;}{a}{}{,}

\defcitealias{Planck2016}{Planck Collaboration XIII}

\begin{document} 

\title{LOFAR discovery and wide-band characterisation of an ultra-steep spectrum AGN radio remnant associated with Abell~1318}

\titlerunning{LOFAR remnant}
\authorrunning{Shulevski et al.}           
\author{Aleksandar Shulevski\inst{1,2,3,13}\orcidlink{0000-0002-1827-0469}
    \and
    Marisa Brienza\inst{7,4}
    \and
    Francesco Massaro\inst{6}
    \and
    Raffaella Morganti\inst{2,3}
    \and
    Huib Intema\inst{1}
    \and
    Tom Oosterloo\inst{2,3}
    \and
    Francesco De Gasperin\inst{5,8}
    \and
    Kamlesh Rajpurohit\inst{4,5,9}
    \and
    Thomas Pasini\inst{8}
    \and
    Alexander Kutkin\inst{2}
    \and
    Dany Vohl\inst{2,13}\orcidlink{0000-0003-1779-4532}
    \and
    Elizabeth A. K. Adams\inst{2,3}\orcidlink{0000-0002-9798-5111}
    \and
    Bj{\"o}rn Adebahr\inst{10}
    \and
    Marcus Br{\"u}ggen\inst{8}
    \and
    Kelley M. Hess\inst{11,2,15}\orcidlink{0000-0001-9662-9089}
    \and
    Marcel G. Loose\inst{2}\orcidlink{0000-0003-4721-747X}
    \and
    Leon C. Oostrum\inst{2,13,14}\orcidlink{0000-0001-8724-8372}
    \and
    Jacob Ziemke\inst{2,12}
        }
        \institute{
        Leiden Observatory, Leiden University, PO Box 9513, NL-2300 RA Leiden, The Netherlands
        \and
    ASTRON, the Netherlands Institute for Radio Astronomy, Postbus 2, 7990 AA, Dwingeloo, The Netherlands
    \email{shulevski@astron.nl}
        \and
        Kapteyn Astronomical Institute, University of Groningen, Landleven 12, 9747 AD Groningen, The Netherlands
        \and
        Dipartimento di Fisica e Astronomia, Università di Bologna, Via P. Gobetti 93/2, I-40129, Bologna, Italy
        \and
        INAF - Istituto di Radio Astronomia, Via P. Gobetti 101, I-40129 Bologna, Italy
        \and
        Dipartimento di Fisica, Università degli Studi di Torino, via Pietro Giuria 1, I-10125 Torino, Italy
        \and
    INAF – Osservatorio di Astrofisica e Scienza dello Spazio di Bologna, Via P. Gobetti 93/3, 40129 Bologna, Italy
    \and
    Hamburger Sternwarte, Universit{\"a}t Hamburg, Gojenbergsweg 112, 21029, Hamburg, Germany
    \and
    Harvard-Smithsonian Center for Astrophysics, 60 Garden Street, Cambridge, MA 02138, USA
    \and
    Astronomisches Institut der Ruhr-Universit{\"a}t Bochum (AIRUB), Universit{\"a}tsstrasse 150, 44780 Bochum, Germany
    \and
    Instituto de Astrofísica de Andalucía (CSIC), Glorieta de la Astronomía s/n, 18008 Granada, Spain
    \and
    University of Oslo Center for Information Technology, P.O. Box 1059, 0316 Oslo, Norway
    \and
    Anton Pannekoek Institute for Astronomy, University of Amsterdam, Postbus 94249, 1090 GE Amsterdam, The Netherlands
    \and
    Netherlands eScience Center, Science Park 402, 1098 XH Amsterdam, The Netherlands
    \and
    Department of Space, Earth and Environment, Chalmers University of Technology, Onsala Space Observatory, 43992 Onsala, Sweden}
\date{\today}
 
  \abstract
  {We present the discovery of a very extended (550 kpc) and low-surface-brightness (3.3 \uJyarcsecsq\ at 144 MHz) radio emission region in Abell~1318. These properties are consistent with its characterisation as an active galactic nucleus (AGN) remnant radio plasma, based on its morphology and radio spectral properties. We performed a broad-band (54 - 1400 MHz) radio spectral index and curvature analysis using LOFAR, uGMRT, and WSRT-APERTIF data. We also derived the radiative age of the detected emission, estimating a maximum age of 250 Myr. The morphology of the source is remarkably intriguing, with two larger, oval-shaped components and a thinner, elongated, and filamentary structure in between, plausibly reminiscent of two aged lobes and a jet. Based on archival {\it Swift} as well as SDSS data we performed an X-ray and optical characterisation of the system, whose virial mass was estimated to be $ \sim 7.4 \times 10^{13} \,$\msun. This places A1318 in the galaxy group regime. Interestingly, the radio source does not have a clear optical counterpart embedded in it, thus, we propose that it is most likely an unusual AGN remnant of previous episode(s) of activity of the AGN hosted by the brightest group galaxy ($ \sim 2.6 \times 10^{12} \,$\msun), which is located at a projected distance of $\sim$170 kpc in the current epoch. This relatively high offset may be a result of IGrM sloshing sourced by a minor merger. The filamentary morphology of the source may suggest that the remnant plasma has been perturbed by the system dynamics, however, only future deeper X-ray observations will be able to address this question.}


\maketitle

\section{Introduction}
\label{intro}

Charged particles moving with relativistic velocities in magnetic fields radiate via the non-thermal, synchrotron radiation mechanism. Astrophysical radio sources are primarily powered in such a way and by studying them in detail, we can determine their physical characteristics, such as the energy density of particles and magnetic field, the time elapsed since the production of the radiating plasma, and so on. With the passage of time, radiative losses mostly affect higher energy particles, resulting in a characteristic steepening of the radio spectra\footnote{$S\sim\nu^{\alpha}$, where S is the flux density, $\nu$ the frequency and $\alpha$ the spectral index}, initially modelled as a power law by \cite{Kardashev1962} and subsequently by spectral curvature models \citep{Tribble1993}. The spectral shape can be used to infer the particle radiative ages and to constrain the acceleration and energising mechanisms, which have given rise to the emission in question \citep[see e.g.][]{Komissarov1994}.

\begin{figure*}[htpb!]
\centering
\includegraphics[width=\textwidth]{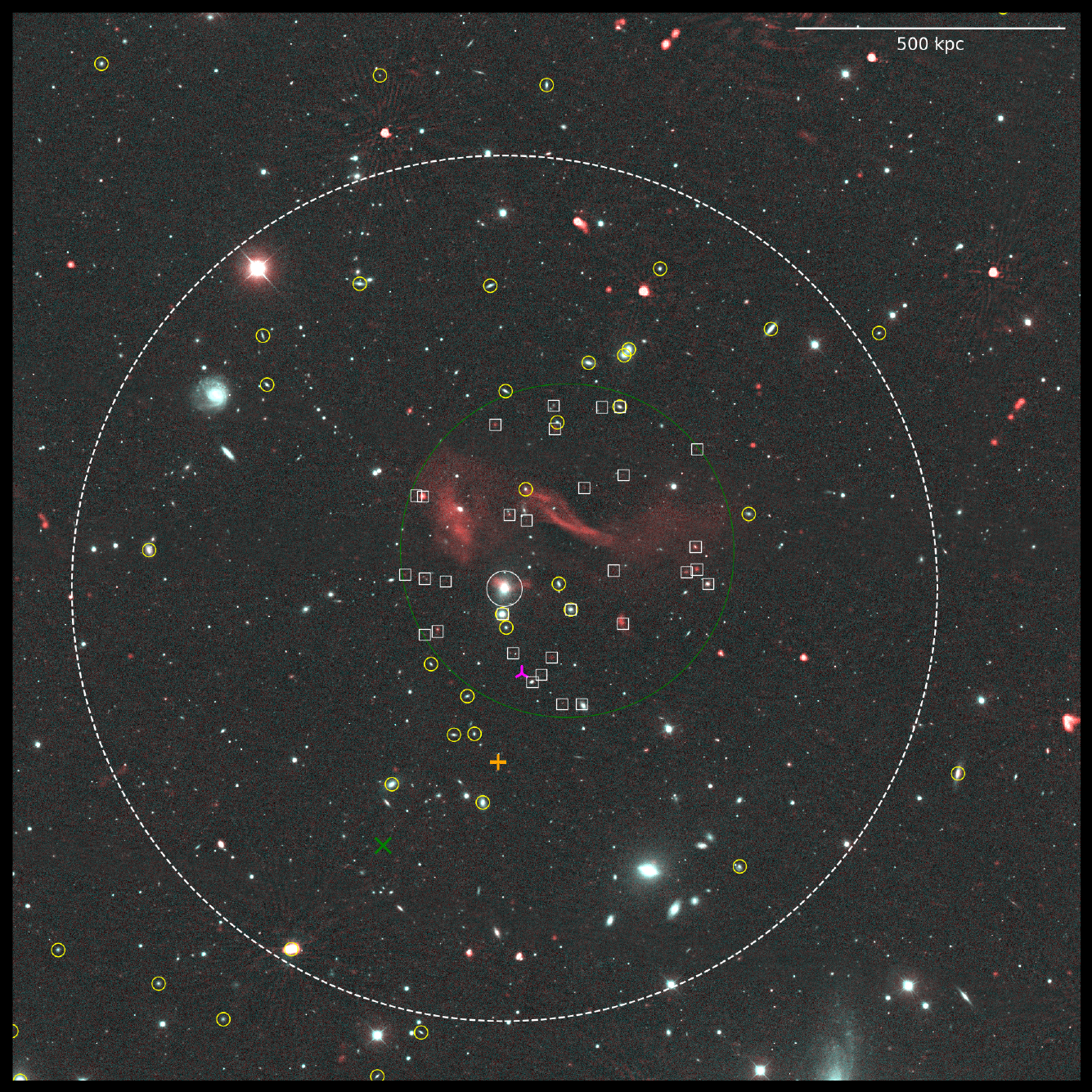}
\caption{LoTSS survey image (red) overlaid on a Panoramic Survey Telescope \& Rapid Response System \citep[Pan-STARRS;][]{Flewelling2020} survey color composite (g, r, and i bands) image. The green cross marks the center position of the galaxy cluster Abell 1318 according to \cite{Abell1989}. The magenta triangle marks the position of an X-ray source from the 2RXS catalog \citep{Boller2016} and the orange plus sign marks the position of the X-ray point source highlighted in Figure \ref{fig:A1318_swift}. The dashed white circle marks the $ R_{200} $ radius of A1318 given in \cite{Wen2012} based on the cluster members found using SDSS DR12 data. The solid green circle marks the aperture in which we have searched for LoTSS catalog entries around the target; the identified sources are marked by white squares and listed in Table \ref{table:A1318_field_sources}. Yellow circles mark the cosmological neighbors of the A1318 BGG (SDSS J113603.51+550430.9) identified by the statistical analysis outlined in Section \ref{env}. The BGG is marked with a solid white circle, while the scalebar in the upper right corner indicates the linear scale for its redshift.}
\label{fig:A1318_composite}
\end{figure*}

Large-scale (up to around Mpc) radio-emitting regions can be related to the activity of active galactic nuclei (AGNs), mainly belonging to the radio-loud class (i.e. radio galaxies and quasars). From the region close to their super-massive black hole, they can eject jets of relativistic particles collimated and accelerated by magnetic fields \citep{Blanford1977, Chatterjee2022}, which then form extended lobes of radio-emitting plasma. The most powerful jetted AGNs are usually hosted by massive elliptical galaxies, as they are often harbored in galaxy clusters or groups, also identified as brightest cluster or group galaxies (BCG/BGG), respectively \citep[see e.g.][]{Best2007}. Once the jet activity ceases, the radio lobes start fading and we observe an AGN radio remnant, often characterised by a prominently curved radio spectrum ($\alpha_{\mathrm{low}} - \alpha_{\mathrm{high}} > 0.5$), its emission getting weaker at higher frequencies as time elapses. These remnant lobes lack compact emission regions, namely,  jets and hotspots, in contrast with active radio galaxies \citep{Parma2007, RefWorks:34, RefWorks:101, Brienza2016, Shulevski2017, Jurlin2021}.

During cluster or group mergers, energy is injected into the intracluster or intragroup medium \citep{Brunetti2014, Vazza2016, Vittor2021}) in the form of turbulence and shocks. This, in turn, can give rise to various kinds of diffuse radio sources, such as radio halos and relics \citep{Weeren2019}. In particular, radio halos are megaparsec-scale sources centered on the cluster core and powered by turbulence \citep{Vazza2011}. Radio relics instead are megaparsec-scale elongated sources located at the outskirts of clusters and trace (merger) shocks \citep{Feretti2012, Stroe2016, Gennaro2021}. In such dynamically disturbed environments, the evolution of AGN remnants can also be affected. For example, in the case of shocks, the AGN remnant plasma can be rejuvenated and start to radiate in the radio domain again \citep{Mandal2019}. Such sources are called radio phoenices to distinguish them from passively fading AGN lobes \citep{Shulevski2015, Gasperin2017}. Moreover, recent simulations clearly show that even minor interactions can easily affect the AGN remnant lobes and spread their content across hundreds of kpc, resulting in complex radio morphologies \citep{Vazza2021, Vazza2023}.

These processes result in radio sources, which due to the nature of the energy input and loss mechanisms, are brighter and preferentially detected at low radio frequencies. Hence, the Low-Frequency Array \citep[LOFAR;][]{RefWorks:157} telescope is an ideal instrument to characterise them. It observes at low frequencies (15 - 190 MHz) and thanks to its design (i.e. providing excellent coverage of short UV spacings), it has very good sensitivity to extended, low-surface-brightness objects. In addition, using its longer baselines, it can achieve high spatial resolution (under 1\arcsec\ at 150 MHz), which is crucial for cross-matching the radio emission to optical data in order to identify the AGN host galaxy or de-blend complex source morphologies. Surveys at low frequencies performed with upgraded, as well as new generation, instruments such as LOFAR and other SKA pathfinders (MWA, MeerKAT, ASKAP) are rapidly adding to our knowledge. More remnant sources were recently discovered \citep{Walker2015, Tamhane2015, Duchesne2019, Randriamanakoto2020, Lal2021}, finally enabling population analysis studies to be carried out \citep{Godfrey2017, Brienza2017, Mahatma2018, Jurlin2021, Quici2021, Dutta2023}.

Here, we report on a discovery of a very faint, diffuse region of radio emission with intriguing morphology in the poorly studied galaxy cluster Abell~1318 at $\mathrm{z=}$ 0.05707 (Figure \ref{fig:A1318_composite}. The source was first identified in the LOFAR Two Meter Sky Survey \citep[LoTSS][]{Shimwell2019, Shimwell2022} at 144 MHz and later followed up at other frequencies to perform a detailed broad-band radio spectral analysis and investigate its nature.

\begin{figure*}[!htpb]
\centering
\includegraphics[width=0.5\textwidth]{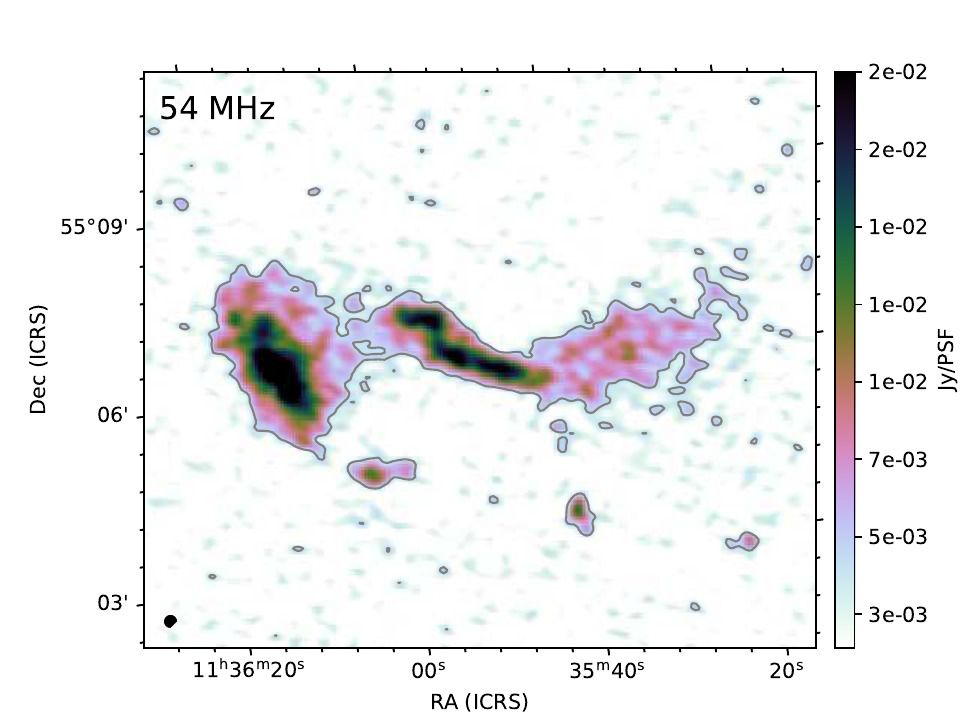}%
\includegraphics[width=0.5\textwidth]{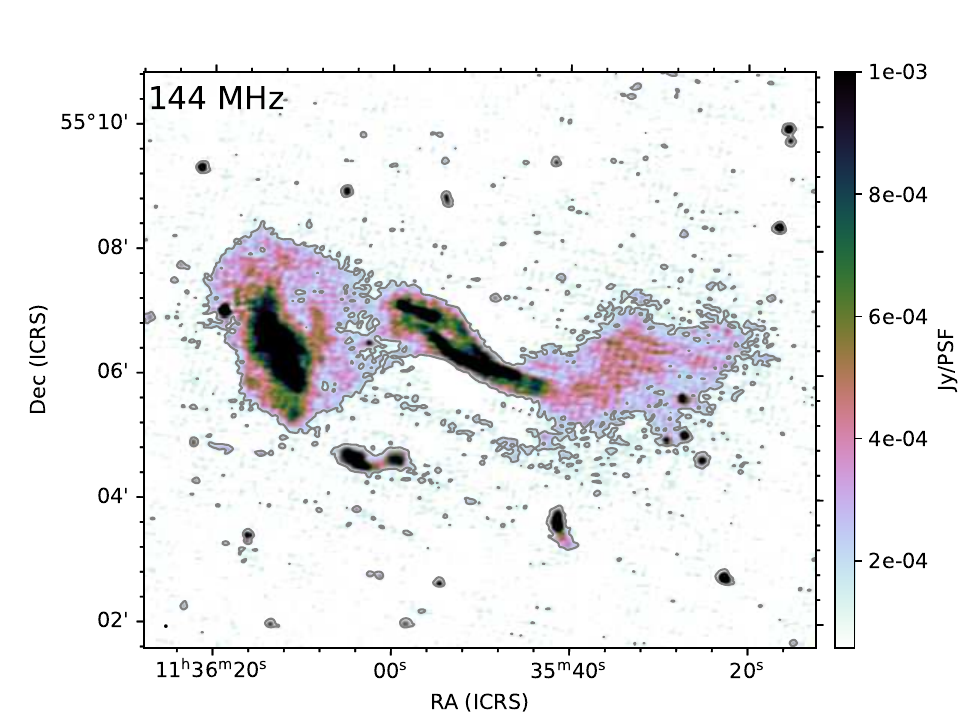}\\
\includegraphics[width=0.5\textwidth]{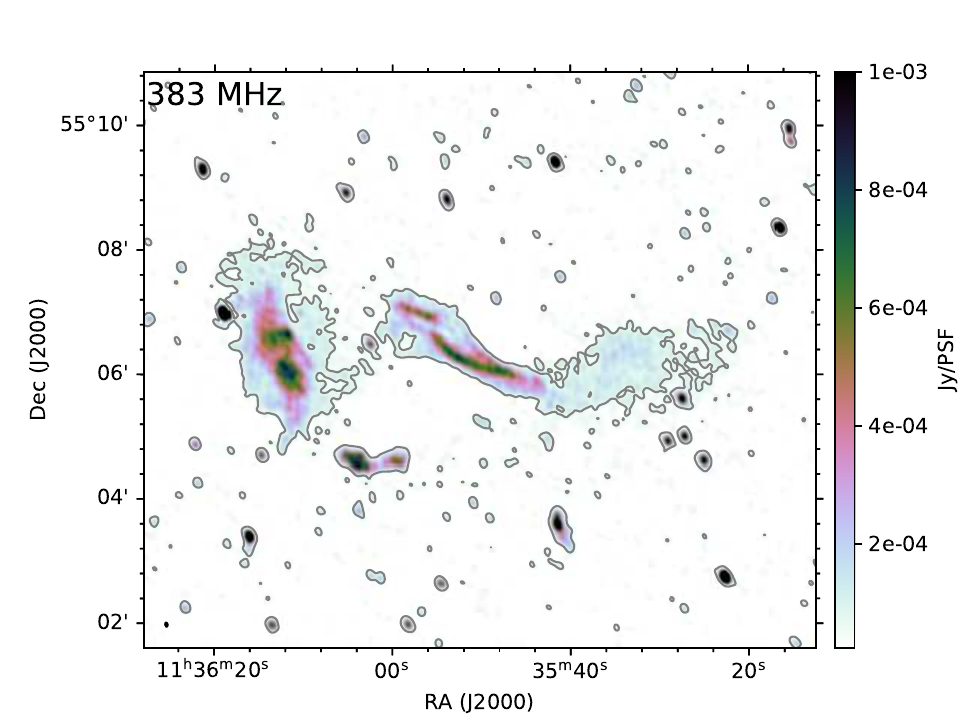}%
\includegraphics[width=0.5\textwidth]{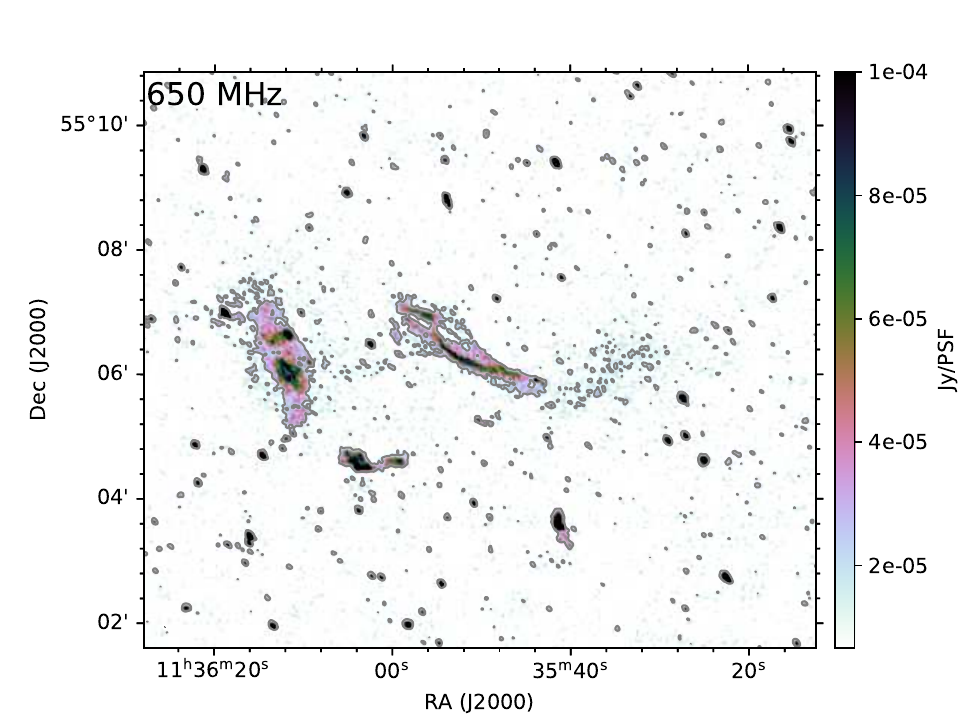}\\
\includegraphics[width=0.5\textwidth]{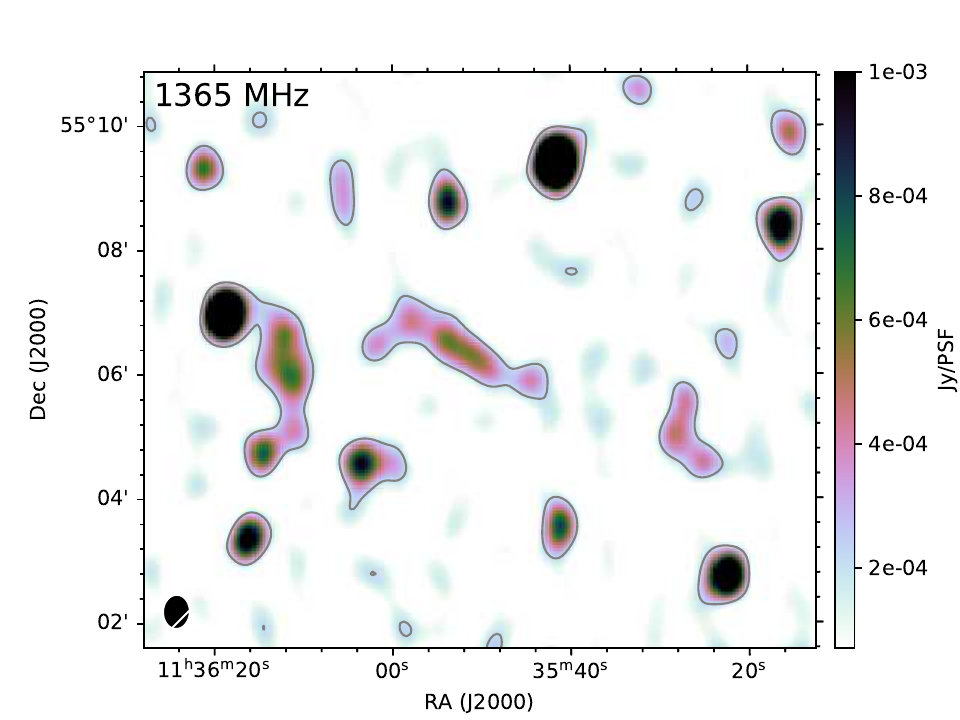}
\caption{Intensity maps of the radio source. Clock\-wise from top left: LOFAR LBA (54 MHz), LOFAR HBA (144 MHz), uGMRT Band 4 (650 MHz), APERTIF (1365 MHz) and uGMRT Band 3 (383 MHz). One contour drawn at 3$\sigma$, image properties listed in Table \ref{table:data}}
\label{fig:A1318_intensity_all}
\end{figure*}

We describe the observing setup and source properties in Section \ref{data}. In Section \ref{res}, we discuss the source energetics, including radio spectral indices between different observing frequencies. We also present a radiative age map based on the radio data and we outline an overview of the X-ray environment the proposed host galaxy is embedded in. Section \ref{dis} elaborates on the results and discusses  future prospects.
We assume a flat $\Lambda$CDM cosmology with $ \mathrm{H}_{0} = 67.8 $\kmsmpc\ and $ \Omega _{\mathrm{m}} = 0.308 $, taken from the cosmological results of the full-mission Planck observations \citep{Planck2016}, unless otherwise stated. Thus, according to these cosmological parameters, 1\arcsec\ corresponds to a distance of 1.14 kpc at $\mathrm{z=}$ 0.05707\footnote{\cite{Wright2006}} which is the Sloan Digital Sky Survey \citep[SDSS DR14;][]{Abolfathi2018} spectroscopic redshift of SDSS~J113603.51+550430.9 (cross ids: CGCG~268-061, CGCG~1133.2+5521, MCG~+09-19-131, WBL~340-004) which is the BGG of Abell~1318.

\section{Data and reduction}
\label{data}

\subsection{Radio}
\label{data:radio}

The source was discovered by inspecting the LoTSS second data release (DR2) image mosaic at 144 MHz of the the Hobby-Eberly Telescope Dark Energy Experiment \citep[HETDEX;][]{Hill2008} field. LoTSS aims to cover the entire northern hemisphere at 6\arcsec\ resolution and $\sim100\rm \ \mu$\Jypsf sensitivity, offering higher sensitivity and resolution than any other low-frequency radio survey in existence. For this work, we adopted LoTSS imaging products. Details on the observing setup, the high-band antenna (HBA) processing pipeline, and available data products are outlined in the LoTSS DR2 paper \citep{Shimwell2022}.

To derive the spectral index properties of the target over a wide frequency range, we also used LOFAR low-band antenna (LBA) images at 54 MHz, part of the LOFAR LBA Sky Survey (LoLSS) first data release \citep[DR1;][]{DeGasperin2023}, reduced using its dedicated processing pipeline \citep[][Pasini in prep.]{Pasini2022, deGasperin2021}, including the w-stacking clean \citep[WSClean;][]{Offringa2014, Offringa2017} imager to obtain the final maps.

Also, we observed the source with the upgraded Giant Metre Wave Radio Telescope \cite[uGMRT;][]{Swarup1991} in band 3 (300-500 MHz), band 4 (550-950 MHz) and band 5 (1050–1450 MHz) on 3, 12, and 18 June 2019 (project ID: 36\_020). Data were processed using the Source Peeling and Atmospheric Modeling pipeline (SPAM,  \citealp{Intema2014}), upgraded for handling new wide-band uGMRT data. Several nearby bright sources were subtracted before the final imaging which was performed using WSClean. We used the full 400 MHz bandwidth for the final image in band 5, however we did not manage to detect the source. Hence, we did not include the band 5 data in our analysis.

We also detected the target in mosaic images from the Westerbork Synthesis Radio Telescope (WSRT) APERture Tile In Focus \citep[APERTIF;][]{Adams2022} survey, performed at 1365 MHz. These were obtained after running the survey calibration and imaging pipeline. They have much lower spatial resolution compared to the LoTSS or uGMRT images, and similar to the LoLSS survey images.

The data and image properties for all of the data sets used are summarised in Table \ref{table:data}. The listed images are shown in Figure \ref{fig:A1318_intensity_all} and images smoothed to a circular PSF\footnote{Throughout this work, we use PSF as a placeholder for the restoring beam of the CLEAN-ed maps used.} used for the analysis in Figure \ref{fig:A1318_chart}.

\begin{table*}[!ht]
\centering
\noindent \caption{Main parameters of the radio images used in this work (see Fig. \ref{fig:A1318_intensity_all}).}
\label{table:data}
\small
\begin{tabular}{c c c c c c c c}
\hline\hline\\
\small Project ID / Survey & Telescope & $\nu_{\mathrm{c}}$ [MHz] & $\Delta \nu$ [MHz] & $\sigma$ [\Jypsf] & PSF [$\arcsec$] & LAS [$\arcmin$] & B$_{\mathrm{max}}$ [$\lambda$] \\
\hline\\
LT16\_004 / LoLSS & LOFAR & 54 & 24 & $1.4\times10^{-3}$ & $15.0$ & 210 & $16.1\times10^{3}$\\
LC3\_008 / LoTSS & LOFAR & 144 & 47 & $57.4\times10^{-6}$ & $6.0$ & 66 & $41.2\times10^{3}$\\
36\_020 & uGMRT & 383 & 200 & $23.1\times10^{-6}$ & $8.7 \times 6.8$ & 32 & $31.9\times10^{3}$\\
36\_020 & uGMRT & 650 & 200 & $6.6\times10^{-6}$ & $4.7 \times 3.4$ & 17 & $54.2\times10^{3}$\\
APERTIF & WSRT & 1365 & 130 & $70.4\times10^{-6}$ & $33.1 \times 30.0$ & 10 & $6.8\times10^{3}$\\
\hline
\end{tabular}
\tablefoot{
(3) Central frequency (4) Bandwidth (5) Image noise (6) Image resolution, arc-seconds (7) Largest angular scale to which the instrument configuration is sensitive (arc-minutes), and (8) Maximum baseline length (wavelengths)}
\end{table*}

\subsection{X-ray}
\label{data:xray}

\begin{figure*}[htp!]
\centering
\includegraphics[width=\textwidth]{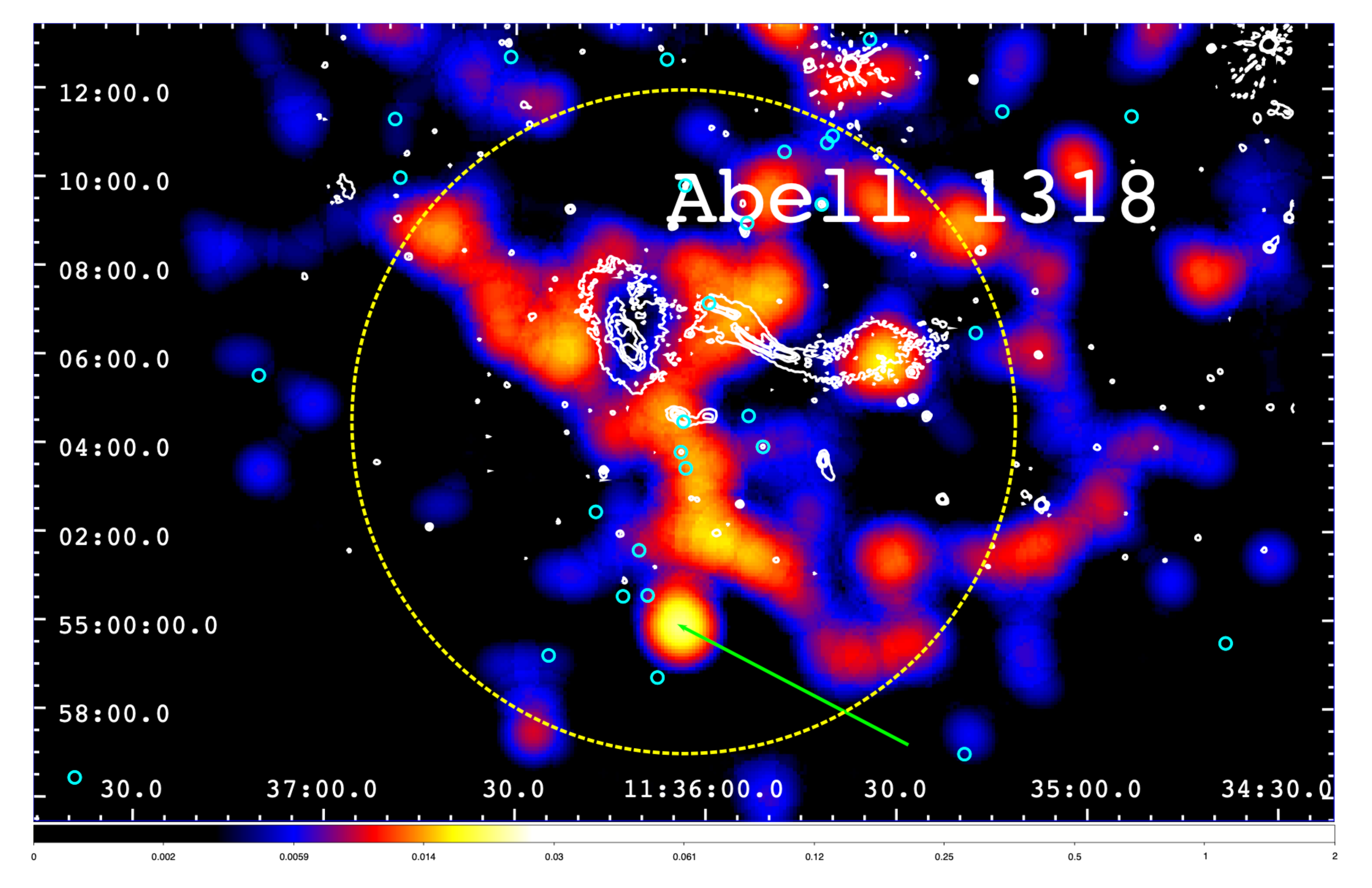}
\caption{{\it Swift}-XRT image of the field around SDSS\,J113603.51+550430.9, obtained from the merged event file and restricted in the 0.5-10 keV energy range. The image is smoothed with a Gaussian kernel of 15 pixels (native pixel size of {\it Swift}-XRT is 2.357\arcsec) and binned by a factor of 2. Radio contours obtained at 6" at 144 MHz are overlaid in white, while cyan circles mark the location of same optical sources highlighted in Figure \ref{fig:A1318_composite}. The dashed yellow circle has a size of 500\,kpc at the redshift of SDSS\,J113603.51+550430.9 while the green arrow indicates the position of a foreground/background source detected in the X-rays.}
\label{fig:A1318_swift}
\end{figure*}

The sky region in the direction of A1318 is covered by the Second ROSAT All-Sky Survey \citep[2RXS;][]{Boller2016} and some tentative emission can be observed associated with the intra-cluster medium. However, a nearby point-like, X-ray source (most likely not associated with A1318), is located in the southern side of the radio relic at an angular separation of $\sim$420\arcsec\ from the radio relic (the magenta cross in Figure \ref{fig:A1318_composite} indicates its optical counterpart, while the green arrow in Figure \ref{fig:A1318_swift} points its X-ray position), thus contaminating the X-ray radiation  potentially arising from the IGM.

We searched several X-ray archives to check for presence of data sets that would allow us to verify the potential X-ray counterpart of radio structures in A1318, as well as emission arising from its intra-group medium. The possible detection of the X-ray emission could, for example, reveal the presence of shocks associated with the gas distribution and/or help confirm the location of the galaxy group center and the BGG since the IGrM typically shows an intensity gradient outward from its centroid \citep[see e.g.][]{Mulchaey2000}. We found two X-ray observations of the field around SDSS J113603.51+550430.9 (within 5\arcmin\ ), with exposure times longer than 1 ksec, in the archive of the X-ray Telescope \citep[XRT;][]{Burrows2000, Burrows2005} on board the Neil Gehrels {\it Swift} Observatory \citep{Gehrels2004}. {\it Swift}-XRT proved to be an extremely valuable X-ray telescope to detect galaxy clusters serendipitously \citep[see e.g.][]{Tundo2012, Liu2013, Liu2015, Dai2015}, thus we reduced and analyzed these X-ray data that could be also useful to plan follow up X-ray observations. The first {\it Swift}-XRT observation was carried out on 21 August  2018 for a total exposure time of $\sim$1.6\,ksec (observation ID: 03105292001), while the second one (observation ID: 00011128001) was performed on August 3, 2019 for $\sim$2\,ksec. 

We processed the datasets using the same data reduction procedures described in previous {\it Swift}-XRT analyses \citep[see e.g.][]{Massaro2008, Massaro2011, Paggi2013, Marchesini2019, Marchesini2020}. We used the \textsc{xrtpipeline} task, part of the {\it Swift}-XRT Data Analysis Software \citep[XRTDAS;][]{Capalbi2005}, developed at the ASI Space Science Data Center and distributed within the HEAsoft package (v. 6.30.1), to obtain clean event files. These were all calibrated and cleaned with standard filtering criteria using the \textsc{xrtpipeline} task (v. 0.13.6), combined with the latest calibration files available in the {\it Swift} calibration database (CALDB) version (v. 20220907) distributed by HEASARC\footnote{\href{https://heasarc.gsfc.nasa.gov/docs/heasarc/caldb/caldb\_supported\_missions.html}{https://heasarc.gsfc.nasa.gov/docs/heasarc/caldb/\\caldb\_supported\_missions.html}}. Events in the energy range of 0.5–10 keV with grades 0–12, in photon counting mode, were used in the analysis \citep[see][for a description of readout modes]{Hill2004}. No signatures of pile-up were found in our {\it Swift-XRT} observations. 

Using \textsc{xselect}, we excluded time intervals having (i) count rates higher than 40 photons per second and (ii) the CCD temperature exceeding -50$^\circ$, in regions at the edges of the XRT camera\citep{Delia2013}. Clean event files were all merged using the \textsc{xselect} task, while corresponding exposure maps were merged with the XIMAGE software and we obtained a merged event file of integrated exposure time $\sim$3.6\,ksec. The final image is shown in Figure \ref{fig:A1318_swift}.

We measured 85$\pm$9 X-ray photons in the 0.5-10 keV energy range within a circular region of 300\arcsec centered at RA 11:36:03.51 DEC +55:04:31 (J2000), on the location of SDSS J113603.51+550430.9, and excluding those within a circular region of 120\arcsec centered on the position of the foreground/background X-ray source marked with the green arrow in Fig. \ref{fig:A1318_swift} This measurement implies a detection significance above 5$\sigma$ level of confidence, assuming a Poissonian distribution of the X-ray background. It is worth noting that the majority of the X-ray photons, namely, 70 out of 85, are detected in the 0.5-3 keV energy range, thus suggesting that the spectrum of the extended emission is rather soft and consistent with being due to ICM emission. This relatively low number of X-ray photons prevents us from carrying out a detailed X-ray spectral analysis. However, it allows us to clearly confirm the presence of extended X-ray emission around the whole radio structure, potentially due to the thermal radiation of the hot intra-group medium. The X-ray diffuse emission is detected in the 0.5-10 keV energy range over a scale of $\sim$400\arcsec, well beyond the {\it Swift}-XRT point spread function, corresponding to $\sim$120\arcsec, according to the latest model released\footnote{\href{https://heasarc.gsfc.nasa.gov/docs/heasarc/caldb/swift/docs/xrt/SWIFT-XRT-CALDB-10\_v01.pdf}{https://heasarc.gsfc.nasa.gov/docs/heasarc/caldb/swift/docs/xrt/SWIFT-XRT-CALDB-10\_v01.pdf}}. Moreover, such X-ray diffuse emission seems to be extended in the NE-SW direction, rather than having a circular shape.

Due to the low number of photons, a detailed morphological investigation of the large-scale X-ray emission around SDSS\,J113603.51+550430.9 is not currently  possible and will require deeper observations. We also note that we did not detect any X-ray emission from SDSS\,J113603.51+550430.9 itself (as point-like source) given the (relatively short) integrated exposure time.

\section{Results}
\label{res}

\subsection{Source properties}
\label{props}

The radio source is located at an angular distance of $ 10.5\arcmin $ (corresponding to a physical scale of $\sim$700 kpc) from the center of the galaxy cluster Abell~1318 reported in \cite{Abell1989} (marked with a green cross in Figure \ref{fig:A1318_composite}).

\begin{figure*}[!htpb]
\centering
\includegraphics[width=\textwidth]{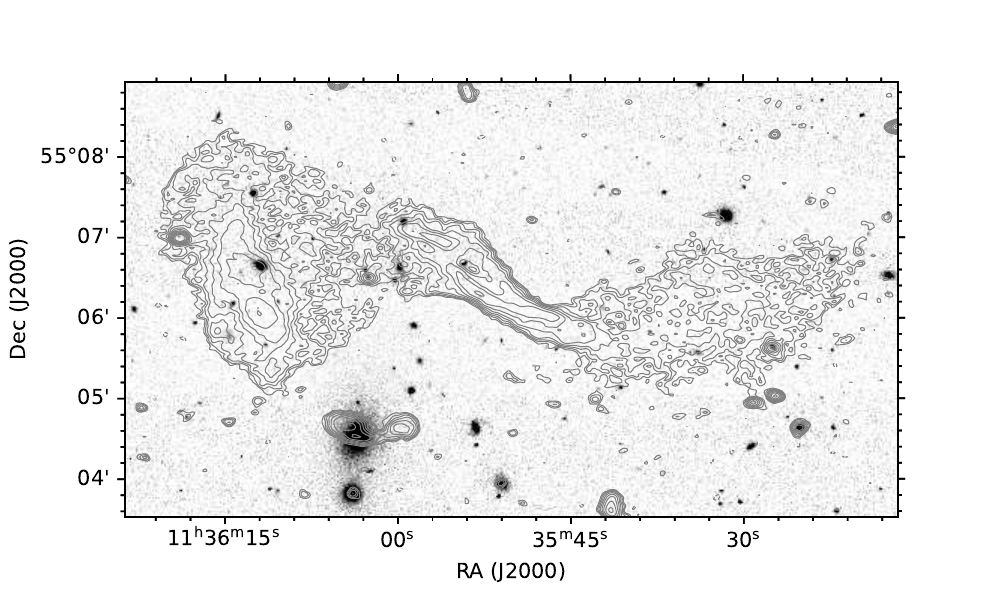}\\
\includegraphics[width=\textwidth]{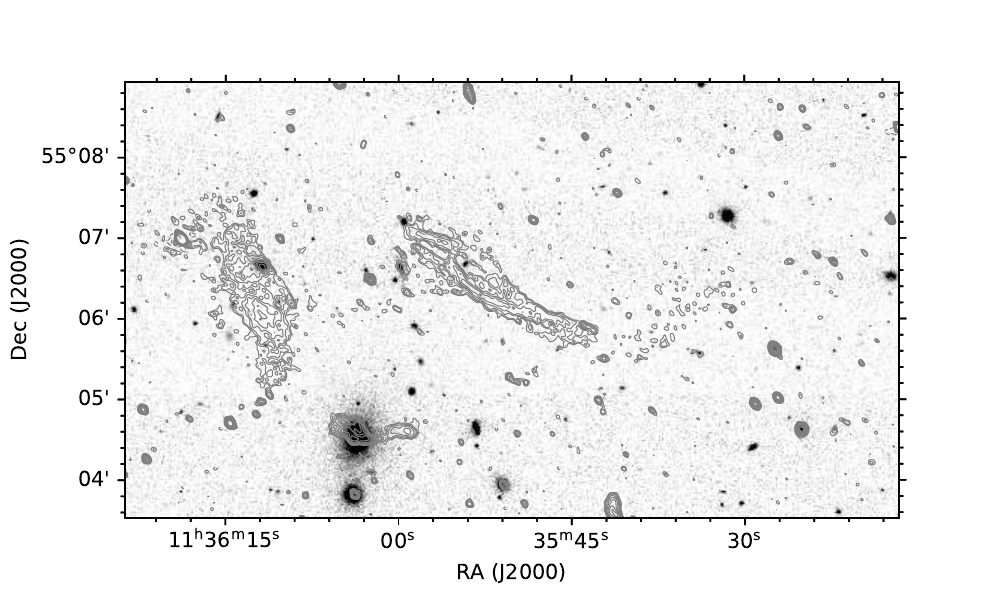}
\caption{PANSTARRS g band optical image centered on the target using an inverted grayscale mapping and linear stretch between levels of $10^{-3}$ and $250 \times 10^{-3}$. Top panel: Ten positive LOFAR radio contours at 144 MHz are overlaid in solid gray lines with levels at $ \left(\sqrt{2}\right)^{n}3\sigma $, where n goes from 0 to 9 in increments of 1. One negative contour level at $ -3\sigma $ is overlaid in a dashed gray line. Bottom panel: Same as in the top panel, using the uGMRT Band 4 image at 650 MHz. The radio images rms noise and resolution are listed in Table \ref{table:data}.}
\label{fig:A1318_radio}
\end{figure*}

\begin{figure*}[!ht]
\centering
\includegraphics[width=0.5\textwidth]{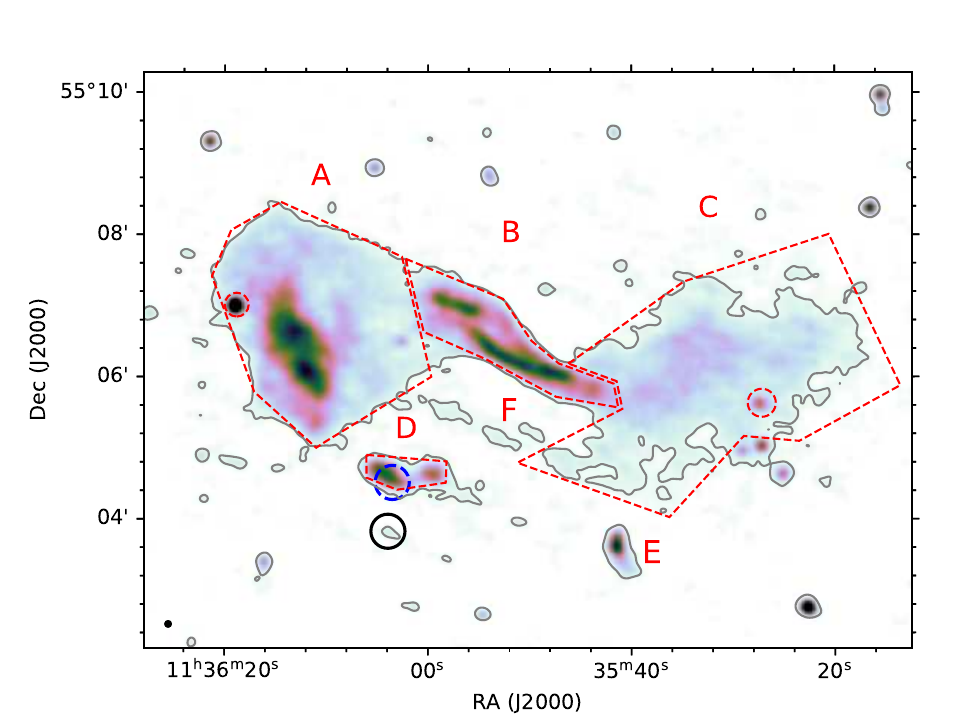}%
\includegraphics[width=0.5\textwidth]{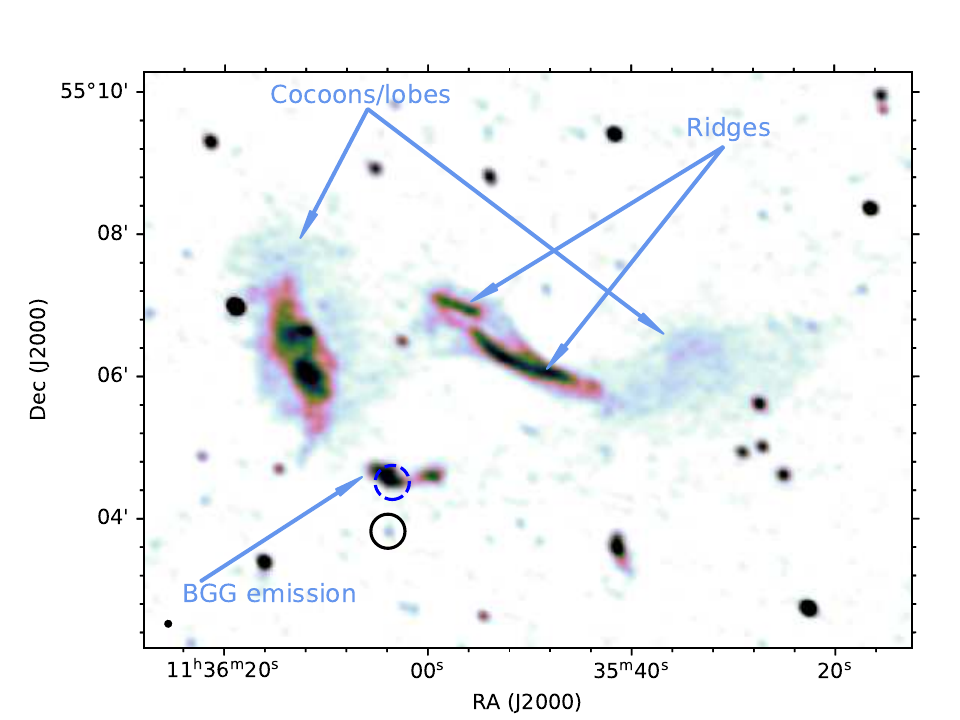}
\caption{LOFAR 144 MHz (left panel) and uGMRT 383 MHz (right panel) intensity maps . The LOFAR map levels go from $\sigma = 88.5 \mu$\Jypsf to 4 \mJypsf, and one contour at a level of $3\sigma$ is overlaid over the LOFAR intensity map. Regions of interest discussed in the text are marked with a red dashed line and labeled; excluded are bright compact sources embedded in the extended emission. The corresponding uGMRT map levels are 10 $\mu$\Jypsf and 1 \mJypsf correspondingly. The PSF size of 9\arcsec\ is shown in the lower left corner. The position of the BGG and its companion are outlined with dashed blue and full black circles respectively and source features of interest are labeled.}
\label{fig:A1318_chart}
\end{figure*}

The radio emission has a total extension of $\sim$8\arcmin, which corresponds to $550$ kpc at the redshift of the system. It has a curious, diffuse, and complex morphology, characterised by two extended components with higher surface brightness (marked A and B in Figure \ref{fig:A1318_chart}). Component A is located to the east of component B, and consists of a region of bright emission, embedded in a low surface brightness (3.3 \uJyarcsecsq $\,\equiv$ 0.3 \mJypsf) cocoon, which narrows as it extends toward component B. Component B has an elongated shape with well-defined edges and encloses a number of peculiar filamentary structures. A shorter filament to the north-east morphs into two filaments to the south west, which then continue into an irregular low-surface-brightness (3.3 \uJyarcsecsq) extension to the west (component C), similar in surface brightness to the cocoon observed in component A. As best highlighted in the uGMRT maps at 383 MHz and 650 MHz (see Figure \ref{fig:A1318_intensity_all} and Figure \ref{fig:A1318_radio}), the filaments are very thin going down to the resolution limit of the images we have, with sizes down to 3.4\arcsec\ at 650 MHz, corresponding to around 4 kpc across. Also, we detect a very faint diffuse emission region to the south of component B, which seems to connect components C and A; we label it F. Overall, the morphology of the source resembles that of a radio galaxy with two lobes of emission connected with filamentary structures. Therefore, it is likely to be the result of AGN activity.

The source has several point radio sources embedded in it, as well as many optical galaxies, as clearly shown in Figure \ref{fig:A1318_radio}. However, none of them seem to plausibly represent a radio core or an obvious optical counterpart for it. There is only one optical galaxy located within the radio emission at RA 11:35:59 DEC +55:07:12 (J2000) and at the same redshift as Abell 1318, not showing any compact radio emission. Most of the radio point sources in the source vicinity have optical counterparts as shown in Figure \ref{fig:A1318_composite}. For example, embedded in the diffuse emission at the eastern edge of component A there is a bright quasar.  

The imaging resolution of LOFAR enables us to perform a robust radio-optical matching. Table \ref{table:A1318_field_sources} lists all the radio sources and their optical matches listed in the LoTSS DR2 catalog contained within a circular aperture centered on RA 11:35:51.57 DEC +55:05:30.9 (J2000) with a radius of $ 4.5\arcmin $ shown in Figure \ref{fig:A1318_composite}. Most of these sources though are not associated with Abell 1318. 

To the south (and unrelated from the main extended radio source), there is also what appears to be a lobed radio galaxy (marked as D, $\sim$170 kpc away in projection) associated with the galaxy host MCG +09-19-131, the system BGG, which does not show any discernible radio core. Interestingly, the optical galaxy host center appears to be shifted with respect to the center-point of the two lobes by around 15\arcsec\ ($17$ kpc). Further to the south-west, there is an unrelated (tailed) radio galaxy (marked as E, $\sim$190 kpc away in projection).
The total (spectral) radio luminosity of the extended radio source (components A+B+C) is $\rm\sim4.6\times10^{24}$\whz, and of component D, associated with the BGG $\rm\sim1.9\times10^{23}$\whz (Table \ref{table:par}). In Figure \ref{fig:A1318_radio} we can also observe that the BGG has a companion elliptical galaxy, located at approximately 48\arcsec\ ($55$ kpc) away towards the south, and showing radio emission as well, corresponding to a luminosity of $\rm\sim3.0\times10^{21}$ \whz at 144 MHz consistent with low power AGN activity (e.g. \citealp{Capetti2022}). In addition, its SDSS optical spectrum clearly resembles that of an elliptical galaxy, lacking [OII], [OIII], and $\mathrm{H}\alpha$ emission lines, supporting the inference that no significant star formation is occurring.

\begin{table*}[!htpb]
\centering
\caption{Measured flux densities at different frequencies for the target source components marked in Figure \ref{fig:A1318_chart} and plotted in Figure \ref{fig:A1318_int_flux}.}
\label{table:flux}
\small
\begin{tabular}{l | c c c c c}
\hline\hline\\
\multirow{2}{*}{\small Component} & \multicolumn{5}{c}{\small S [mJy] at $\nu$ [MHz]} \\
\cline{2-6}\\
 & 54 & 144 & 383 & 650 & 1365\\
\hline\\
A & $ 764.39 \pm 9.57 $ & $ 256.68 \pm 0.68 $ & $ 50.78 \pm 0.38 $  & $ 17.62 \pm 0.20 $ & $ 3.42 \pm 0.50 $\\
B & $ 322.93 \pm 12.36 $ & $ 120.27 \pm 0.77 $ &  $ 26.85 \pm 0.41 $  & $ 9.44 \pm 0.14 $ & $ 2.28 \pm 0.34 $\\
C & $ 493.46 \pm 4.27 $ & $ 175.03 \pm 0.23 $ & $ 24.15 \pm 0.09 $ & $ 8.1 \pm 0.04 $ & $ 0.66 \pm 0.10 $\\
D & $ 40.55 \pm 7.26 $ & $ 22.89 \pm 0.72 $ & $ 7.08 \pm 0.52 $ & $ 2.96 \pm 0.22 $ & $ 0.80 \pm 0.56 $\\
\hline
\end{tabular}
\end{table*}

\begin{figure*}[!htpb]
\centering
\includegraphics[width=0.5\textwidth]{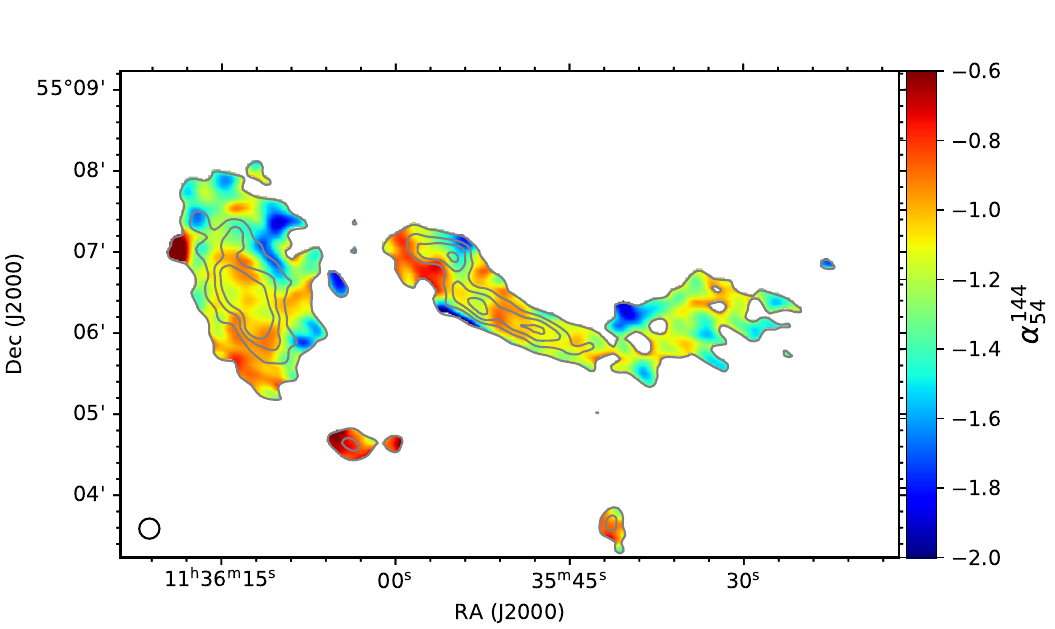}%
\includegraphics[width=0.5\textwidth]{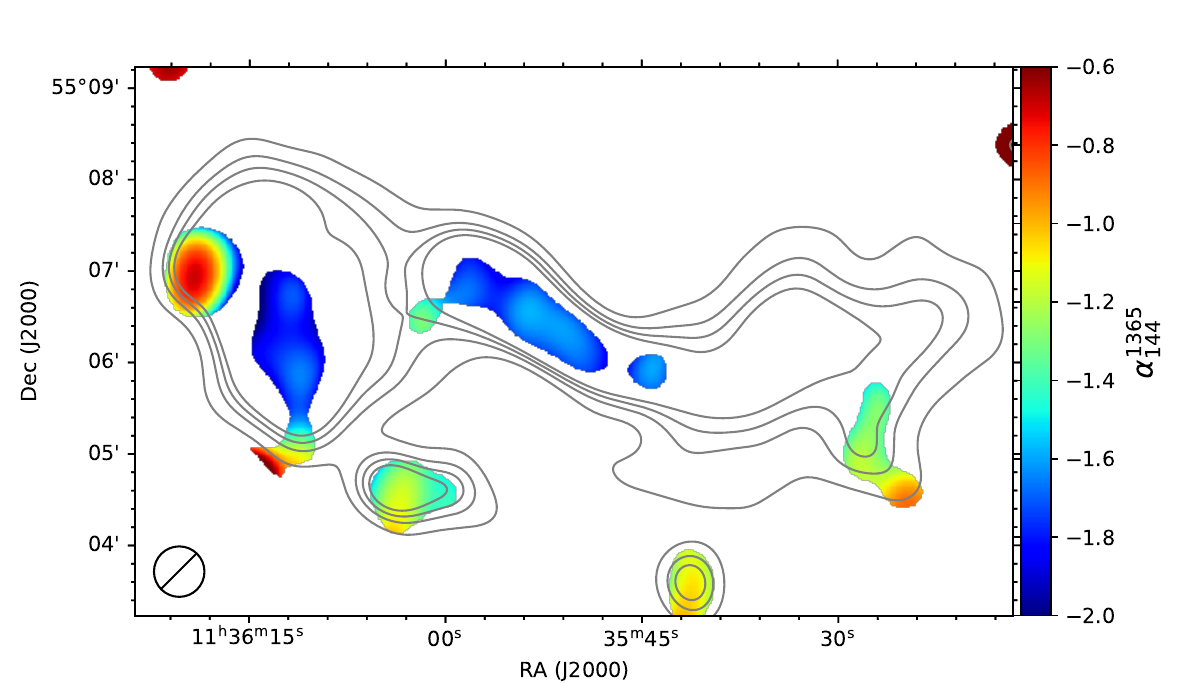}\\
\includegraphics[width=0.5\textwidth]{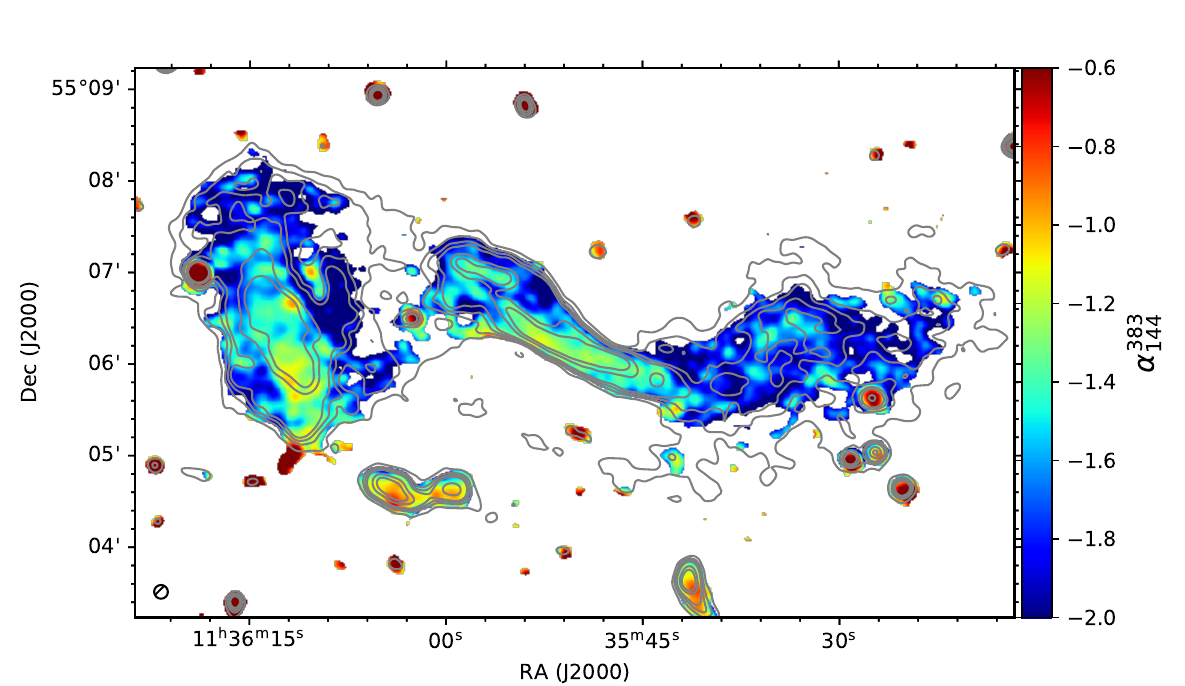}%
\includegraphics[width=0.5\textwidth]{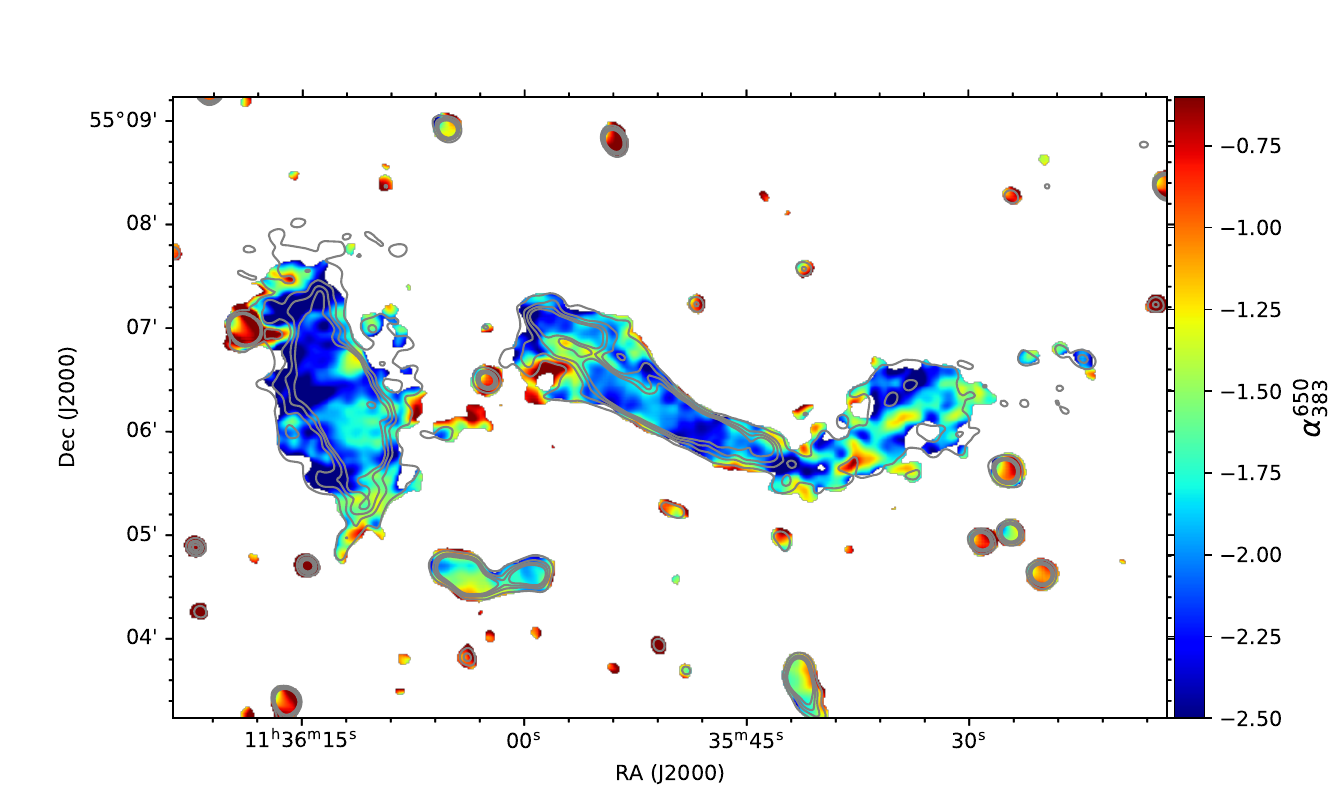}
\caption{Spectral index maps between different frequencies. Top-left panel: $\alpha_{54}^{144}$ (LOFAR/LBA - LOFAR/HBA) spectral index map. Shown are LBA contours at levels of [-5, 5, 9, 12, 15]$\sigma$ and only pixels with values above 5$\sigma$ in both maps are taken into account.  LBA: $\sigma=1.32\,$\mJypsf, HBA: $\sigma=0.13\,$\mJypsf. Assumed flux scale accuracy is 20\%. PSF size: 15$\arcsec$. Top-right panel: $\alpha_{144}^{1365}$ (LOFAR/HBA - APERTIF) spectral index map for pixels with values above 3$\sigma$. Shown are HBA contours at [-5, 5, 9, 12, 15]$\sigma$. HBA: $\sigma=0.6$\mJypsf, WSRT: $\sigma=0.1$\mJypsf. Assumed flux scale accuracy: HBA - 20\%, WSRT - 5\%. PSF: 33$\arcsec$. Bottom-left panel: $\alpha_{144}^{383}$ (LOFAR/HBA - GMRT) spectral index map. Shown are HBA contours at [-5, 5, 9, 12, 15, 27, 36.]$\sigma$. for pixels with values above 3 $\sigma$ in both maps. Bottom-right panel: $\alpha_{383}^{650}$ (GMRT) spectral index map for pixels above 3$\sigma$. Shown are GMRT 385 MHz contours at [-5, 5, 9, 12, 15]$\sigma$. HBA: $\sigma=64.5 \, \mu$\Jypsf, GMRT: $\sigma=26.7 \, \mu$\Jypsf and $15 \, \mu$\Jypsf. Assumed flux scale accuracy: HBA - 20\%, GMRT - 5\%. PSF size: 9$\arcsec$. The corresponding error maps are shown in Figure \ref{fig:A1318_spixerr-LBA-HBA-GMRT-WSRT33_47asec} in the appendix.}
\label{fig:A1318_spix-LBA-HBA-GMRT-WSRT33_47asec}
\end{figure*}

\subsection{Spectral index}
\label{spix}

\begin{figure}[!htpb]
\centering
\includegraphics[width=0.5\textwidth]{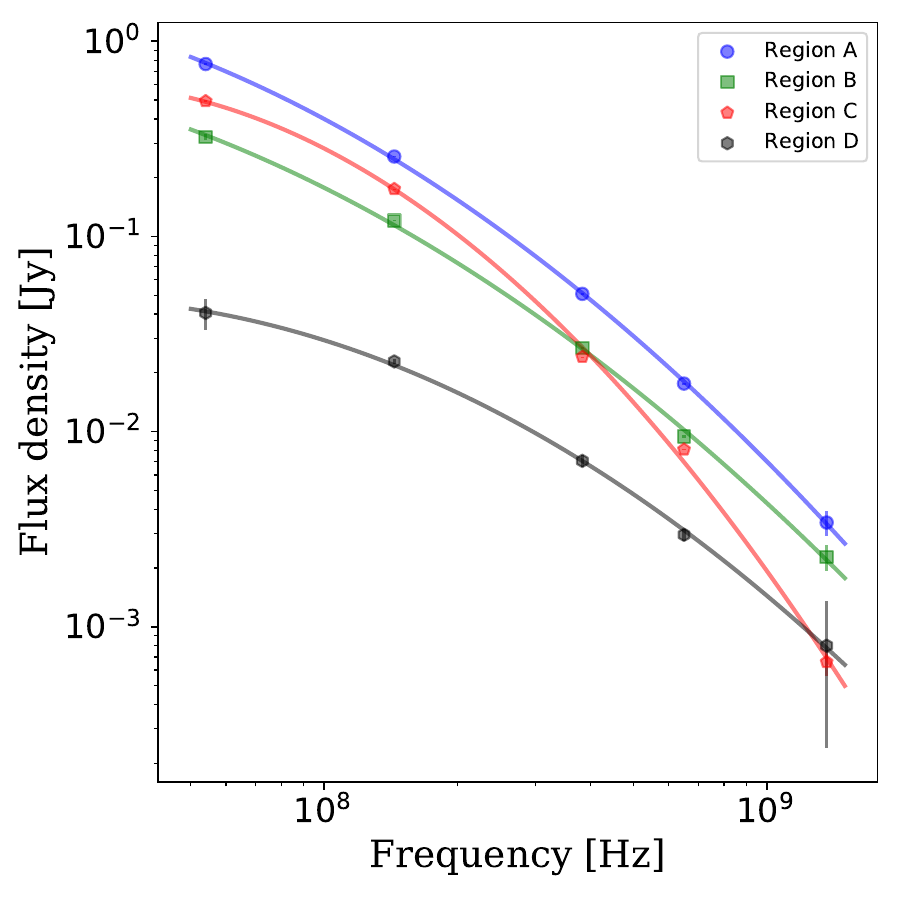}
\caption{Integrated flux densities of the source components listed in Table \ref{table:flux}. Lines show second-order polynomial fits to the data points to highlight the spectrum shape.}
\label{fig:A1318_int_flux}
\end{figure}

The integrated radio spectrum of the different components of the source across the full range of frequencies (54 - 1365 MHz) is very curved, as it is shown in Figure \ref{fig:A1318_int_flux}. The flux densities used in the plot are reported in Table \ref{table:flux} and were measured off of the images shown in Figure \ref{fig:A1318_intensity_all} using the polygons marked with red dashed lines shown in Figure \ref{fig:A1318_radio}.

To explore this trend further, we derived spectral index maps of the source across the full band (see Fig. \ref{fig:A1318_spix-LBA-HBA-GMRT-WSRT33_47asec}). We also created color-color plots and mapped its radiative age. To that end, we smoothed all the input images in the set to the resolution of the lowest frequency one using the {\tt CASA} \citep{McMullin2007} task {\tt imsmooth}, then registered them using one image as a common template with the {\tt imregrid} task. We also verified that after registering the images in a given set do not have appreciable sub-pixel position shifts among them. Only pixels above 3$\sigma$ in all maps were considered in the spectral analysis. The corresponding spectral index error maps are presented in Appendix \ref{app:err}. The spectral index was computed as $ \alpha = \frac{\log(S_{1} / S_{2})}{\log(\nu_{1} / \nu_{2})} $, while the spectral index error according to $ \Delta \alpha = \frac{1}{\ln (\nu_{1} / \nu_{2})} \sqrt{\left( \frac{\sigma_{1}}{S_{1}}\right)^{2} + \left(\frac{\sigma_{2}}{S_{2}}\right)^{2}} $. 

The top-left panel in Figure \ref{fig:A1318_spix-LBA-HBA-GMRT-WSRT33_47asec} shows the derived spectral index map between the LOFAR 54 MHz and 144 MHz images. The resulting map has 15\arcsec resolution, which corresponds to the highest resolution of the LBA image, and it shows clear systematic flattening of the spectral index in components A, B, and C to around $\rm\alpha=$-0.8, especially over the southern sections of component A, closer to component D. Component D itself has the flattest spectral index of around $\rm\alpha=$-0.7

In the top-right panel, we present the 144 MHz - 1365 MHz spectral index map at the resolution of the APERTIF image (33\arcsec) with overlaid contours according to the lower frequency image. Disregarding background sources with flat spectral indices ($\rm\alpha \sim$ -0.6), we detected two areas of interest, associated with the highest surface brightness areas within components A and B, having a steep spectral index of $\rm\alpha=$ -1.5 to $\rm\alpha=$ -1.8. We also detect component D (barely noticeable as an extension to component A in the map discussed above). We stress that despite including higher frequency data, the spectral index observed between LOFAR-APERTIF is flatter than what is observed with the uGMRT maps between 383-650 MHz, as it includes a larger frequency baseline. Due to the high noise level of the APERTIF image no useful spectral index limit can be obtained for the source components where no emission is detected.

In the bottom-left panel of Figure \ref{fig:A1318_spix-LBA-HBA-GMRT-WSRT33_47asec}, we show the LOFAR/HBA - uGMRT spectral index map between 144 - 383 MHz and in the bottom-right panel,  the uGMRT 383 - 650 MHz spectral index map. Both have a PSF size of 9\arcsec\ and they map  most of the source components seen at 144 MHz reasonably well. They even show, via the flatter ($\rm\alpha=$ -0.6) spectral index of the point source associated with it, that the companion galaxy of the BGG appears to host an AGN active at the present epoch. 

In the frequency range of 144-383 MHz, the diffuse emission has a very steep spectral index up to $\rm\alpha=$ -2, which is consistent with aged plasma. The spectrum steepens further in the frequency range 383-650 MHz. The filaments in component B and a southern patch of component A clearly have  a flatter spectral index of around $\rm\alpha_{144}^{383}=$ -1.5. Component D shows a spectral index of around -1.2; however, we note that there is an indication of a somewhat flatter spectral index emission ($\rm\alpha=$ -0.8) in correspondence to the larger elliptical galaxy, which is slightly offset to the east with respect to the center-point between the two lobes of component D.

Taken together, the spectral index mapping shows that the components of the source in which the radiating particles have undergone the most energy loss are the northern parts of components A and B, and especially component C. This discrepancy can possibly be accounted for by the observed morphology: components A and B seem to be more confined, indicating that adiabatic energy loss may prevail in component C.

\begin{figure*}[!]
\centering
\includegraphics[width=\textwidth]{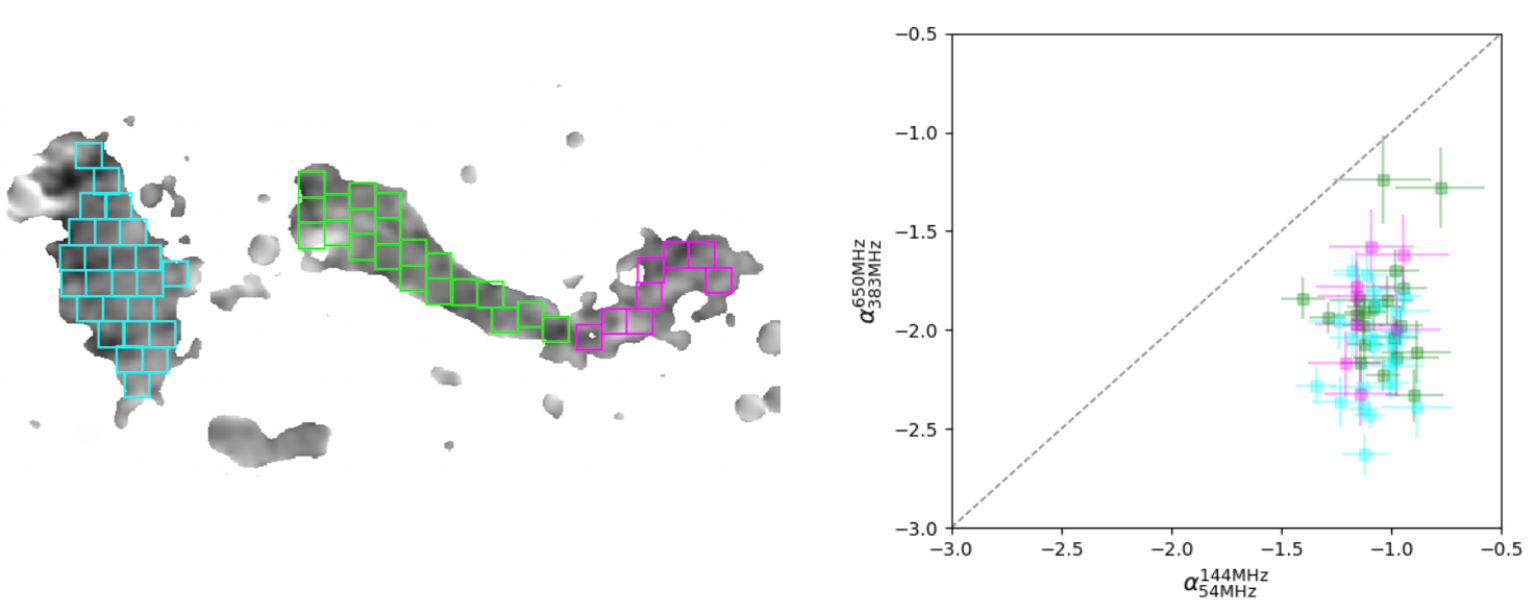}\\
\caption{Color-color plot for different regions in the source obtained using images at 54 MHz, 144 MHz, 383 MHz and 650 MHz with a PSF size of 15\arcsec. The gray dashed line represents the place where $\rm \alpha_{54MHz}^{144MHz}=\alpha_{383MHz}^{650MHz}$. The regions used to extract the values are shown in the left panel, overlaid on the spectral index map in the range 383-650 MHz.}
\label{fig:A1318_cc-plot}
\end{figure*}

To further investigate the spectral curvature throughout the source we also performed a color-color analysis \citep{Katz1993, Katz1997} as shown in Figure \ref{fig:A1318_cc-plot}. For this we used the four images at the lowest frequencies with a PSF of 15\arcsec: 54 MHz, 144 MHz, 383 MHz, and 650 MHz. The gray dashed line in the plot represents the region where $\rm \alpha_{low}=\alpha_{high}$ (i.e. the power-law line). The regions used to perform the measurements have the same size as the PSF and are shown in the left panel of the figure, color-coded according to their position. As we can see from plot, most points lie below the bisector line, in agreement with an aged, curved radio spectrum in all components: A, B, and C \citep[see e.g.][]{Brienza2020} as well as consistent with the inferences drawn from the spectral index mapping. 

\subsection{Spectral age estimates}
\label{spage}

Finally, we derived the spectral age of the plasma across the source, based on the assumption of purely radiative energy loss. We used the {\it PySynch}\footnote{\href{https://github.com/mhardcastle/pysynch}{https://github.com/mhardcastle/pysynch}} python package for calculating the source properties needed as input for the age analysis. We calculated the values of the magnetic field, particle energy density, and pressure for each of the source components shown in Figure \ref{fig:A1318_chart}, assuming each component is circumscribed with cylinder of a given radius and height. The adopted parameters and obtained results are reported in Table \ref{table:par}.

\begin{table}[!ht]
\begin{threeparttable}[b]
\centering
\noindent \caption{Measured and derived physical parameters for the components shown in Figure \ref{fig:A1318_chart}. The flux density values were measured using the 144 MHz map.}
\label{table:par}
\small
\begin{tabular}{l c c c c c}
\hline\hline\\
\small ID & $ l \times h $ [\arcsec] & L$_{\mathrm{R}}$[10$^{24}$\whz] & B [nT] \tnote{*}& e [fJ/m$^{3}$] & p [nPa]\\
\hline\\
A & 210 $\times$ 77 & 2.13 & 0.10 & 7.82 & 330.54\\
B & 217 $\times$ 25 & 1.00 & 0.15 & 18.70 & 500.44\\
C & 290 $\times$ 77 & 1.45 & 0.08 & 5.23 & 270.27\\
D & 67 $\times$ 13 & 0.19 & 0.19 & 28.81 & 634.25\\
\hline
\end{tabular}
\begin{tablenotes}
\item [*] 1T = $10^{4}$G
\end{tablenotes}
\tablefoot{
(1) Component ID; (2) Radius and height of the cylinder circumscribing the component; (3) Computed radio luminosity; (4) Magnetic field in nano-Tesla; (5) Derived particle energy density; and (6) Derived pressure
}
\end{threeparttable}
\end{table}

\begin{figure}[!htpb]
\centering
\includegraphics[width=0.5\textwidth]{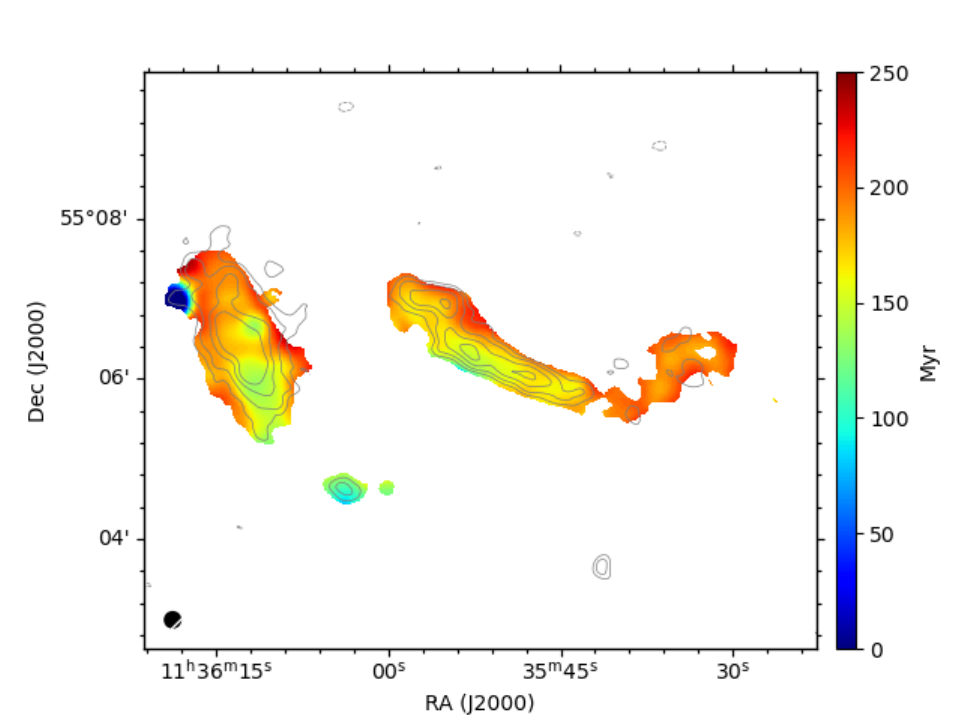}
\caption{Spectral age map of the target based on a JP model fit. 54 MHz map contours are overlaid, and the color indicates the associated age in Myr. Older regions have redder colors.}
\label{fig:A1318_age_map}
\end{figure}

We then used the {\it BRATS} package \citep{Harwood2013, Harwood2015} to estimate the radiative spectral age of the target source. For the magnetic field across the source, we took the average of the values found for the different components ($0.13$ nT) and we searched for the best fit injection spectral index using the \textsc{findinject} task. It fits spectral ageing models by varying the injection index over a fixed range. The minimum of the goodness of fit gives the injection index that we use in the subsequent analysis. The injection index best fit to the data was found to be $\rm\alpha=$ -0.72. The data set used for this operation consisted of the LOFAR (LBA + HBA) and uGMRT (band 3 and band 4) data smoothed to the 54 MHz PSF size.
Taking the above values for the magnetic field and injection index, we  fit a Jaffe-Perola \citep[JP;][]{Jaffe1973} ageing model to the data set (using the \textsc{fitjpmodel} task) thus deriving a spectral age map of the source, which is shown in Figure \ref{fig:A1318_age_map}. We used the JP model since it is the simplest one describing the physics when modeling old remnants; the activity episode is assumed to be short, compared to the plasma age, and the pitch angle of the radiating particles relative to the magnetic field direction is assumed to be isotropised on timescales shorter than the radiation timescale.

Overall, the age of the plasma ranges between 100 and 250 Myr (discounting the area around the embedded background QSO on the western edge), with the diffuse components A and C having the highest age values. The youngest plasma is found in component D, especially towards the BGG, with values of around 100 Myr. We remark that radiative ages derived here are based on the assumption that radiative cooling is the dominant process for particle energy losses, adiabatic energy loss (if any) is ignored, and no re-acceleration is occurring.

\subsection{Large-scale environment analysis}
\label{env}

To better understand the nature of the radio source, we investigated its surrounding environment in optical wavelengths in addition to radio.
We considered the center position of A1318 as reported in the Abell catalog \citep{Abell1989} indicated at a redshift of $z=0.0566$ and then we selected the brightest SDSS sources (brighter than $ m_{\mathrm{R}} $ < 17.8) in the R band within a circular region of 500\arcsec\ from this position identified with SDSS\,J113603.51+550430.9 (MCG +09-19-131), located at a redshift of $z=0.05707$. 

Using this optical source as a proxy, we computed all environmental parameters according to the procedure \citep{Massaro2019, Massaro2020b, Massaro2020a} and \cite{Capetti2020}. We found that the number of so-called cosmological neighbors, defined as all optical sources with SDSS magnitude flags indicating a galaxy-type object and having a spectroscopic z with $\delta\,z =\, \mid z_{src} - z \mid \, <0.005$, lying within 500\,kpc, 1\,Mpc and 2\,Mpc is 18, 36, and 54, respectively. SDSS J113603.51+550430.9 is actually the brightest galaxy lying at a projected distance of 34 kpc from the centroid of the positions of all cosmological neighbors within 2\,Mpc, computed at its source redshift. 
The redshift dispersion of all cosmological neighbors allowed us also to compute the velocity dispersion of the system, which is found to be $\sigma_\mathrm{v}\sim400\,\mathrm{km/s}$ corresponding to a virial mass of $M_{\mathrm{vir}}\sim \frac{\mathrm{R}_{\mathrm{vir}}\,\sigma_\mathrm{v}^{2}}{\mathrm{G}}\sim7.4\times10^{13}\,\mathrm{M}_{\mathrm{sun}}$, assuming a virial radius $\mathrm{R}_{\mathrm{vir}}=2\,\mathrm{Mpc}$. Thus, using the $\mathrm{L_X}\,-\,\sigma_\mathrm{v}$ correlation \citep{Popesso2005} we derived the result that the expected X-ray luminosity in the 0.1-2.4 keV range for the galaxy cluster is $\mathrm{L_X}=0.2-6\times10^{36}\,\mathrm{W}$, when considering the uncertainties on the correlation parameters. Then, adopting a luminosity distance of 256.6\,Mpc and a Galactic column density $\mathrm{N_H}=8.35\times10^{19}\,\mathrm{cm}^{-2}$ \citep{Kalberla2005}, we  converted the {\it Swift}-XRT count rate, measured in 0.5-2 keV energy range, into an X-ray luminosity of $\mathrm{L_X}=5.3\pm{0.2}\times10^{35}\,\mathrm{W}$ in the 0.1-2.4 keV energy range, assuming a thermal model with a temperature of $\mathrm{T}=2\,\mathrm{keV}$. This is consistent with the expected value within a 3$\sigma$ level of confidence. We conclude that Abell 1318 can be described as a relatively high mass galaxy group rather than a cluster.

We also note that the magnitude difference between the BGG and its companion elliptical galaxy in the south is only $\Delta\rm m$=0.64, which is unexpected for a relaxed system dominated by a single massive galaxy. The mass of the two galaxies (as computed from the 2MASS k-band magnitude and using the relation found by \citealp{Cappellari2013}) is log$M_{\sun}$=11.91 and log$M_{\sun}$=11.98, respectively.

This, together with the spatial distribution of the system member galaxies, which tentatively appears to be elongated in north-south direction, and the complex morphology of the X-ray emission as detected by {\it Swift}, suggests that the system is likely not in a relaxed state.

Overall, our statistical analysis certainly confirms the location of the group centroid from an optical perspective. However, to obtain a full overview of the system, it will be necessary to compare it with deeper X-ray observations.

\section{Discussion and conclusions}
\label{dis}

Based on the analysis presented here, we interpret that the observed extended radio source is an old AGN remnant. The morphology of the radio emission, relative vicinity of a plausible AGN host, and the group environment support this conclusion, in addition to the spectral index and spectral age properties in relation to the source structure. In this scenario, components A and C would represent two fading lobes, and component B would represent the old plasma associated with a past jet.
However, the identification of an optical host remains challenging. As we discuss in Section \ref{props}, we did not find any association of flatter spectrum patches (indicating a radio core) within the radio source with any optical galaxies.

One possibility is that the host is the galaxy at RA 11:35:59 DEC +55:07:12 (J2000), which is the only member of Abell 1318 located within the extended radio emission (within component B). However, the galaxy is not very massive (log$M_{\sun}$=10.8 based on the 2MASS magnitude $M_K$=-23.55 and using the relation in \citealp{Cappellari2013}) making it less likely associated with such an extended radio AGN. Moreover, it does not show any compact radio emission and it is very much offset with respect to the radio source barycenter. Also, large-size radio galaxies in clusters/groups are generally hosted in elliptical galaxies \citep[see e.g.][and references therein]{Banfield2015}.

All other radio point sources embedded in the extended emission, or in its immediate vicinity, are background objects and, thus, not associated with it. One interesting possibility that we would like to put forward is that the old radio plasma was originally associated with the system's BGG or its companion to the south.

A curious point, which also hints at our hypothesis, is that the spectral index of component D is, on average, flatter (indicating a particle population that was energised more recently) compared to the larger emission to the north (components A+B+C). It also contains a distinct flatter-spectrum ($\alpha\sim - 0.8$; see bottom panels of Figure \ref{fig:A1318_spix-LBA-HBA-GMRT-WSRT33_47asec}) patch of emission spatially associated with the position of the BGG, which could be evidence of the last parts of this component of the source to be energised as the AGN activity was terminating.
Component D could then represent a second, more recent phase of activity of the AGN hosted by the BGG. The radio emission of component D is slightly offset from the galaxy barycenter, its spectrum reaches very steep values at high frequencies (up to -1.8), and there is no evidence of a radio core. All this suggests that, similarly to the components of the source to the north it is also an AGN remnant. 

If this interpretation is correct, it would mean that the BGG was active around 250 Myr ago  (its activity lasting around 80 Myr based on the age gradient across source components A, B, and C) and it had another activity episode around 100 Myr ago (Figure \ref{fig:A1318_age_map}), of which component D is the remnant. We note that these numbers are valid under the assumption the plasma is evolving passively, namely, in the absence of particle compression or re-acceleration, which is possibly not the case in this system, as further discussed below.

Notably, the BGG companion galaxy to the south shows some compact radio emission as well (noticeable in the bottom panels of Figure \ref{fig:A1318_spix-LBA-HBA-GMRT-WSRT33_47asec}) and it could, in principle, also be considered a potential host of the large-scale emission (components A+B+C) and of component D as well. If so, we would be observing a third epoch of activity for this source. However, this would require a very precise spatial alignment between the trajectory of motion and remnant or active regions; hence, we favor the proposed remnant(s) association with the BGG rather than with its companion. We note that both the BGG and its companion show SDSS spectra devoid of emission lines, hinting that there is no AGN activity at the present epoch.

The fact that both remnants are offset from the current position of the BGG can be explained by the movement of the BGG through the intra-group medium (IGrM) of Abell~1318.
This appears consistent with the idea of a dynamically active system, as tentatively suggested by both its observed X-ray (complex morphology) and optical properties (see Sect. \ref{env}). In this context, the BGG would be still moving with a substantial velocity in the group gravitational potential and the relatively large distance (in projection) between it and the radio remnant(s) is not all that unexpected.
If we assume an average speed equal to 400 km/s (equal to the dispersion velocity of the system) and a distance (in projection) between the BGG and the center of component B of 2\arcmin ($\sim$140 kpc) we get, as a lower limit, that it takes $\sim$340 Myr to displace the BGG from region B to its present position. 

We note that this timescale is larger than the minimum value of the radiative age computed throughout the source equal to $\sim$150 Myr, which should represent the age of the particles last accelerated by the jet. This inconsistency could partially be reconciled if the true BGG velocity was higher than what assumed. However, to match the timescales, a higher velocity by a factor of 2 would be needed, which is probably too high for this kind of system. At the same time if the system is dynamically active, it is reasonable to believe that the AGN remnant plasma itself (component A+B+C) might have been transported away by the IGrM motions, even in opposite direction with respect to the BGG movement direction. This occurrence, for example, has been clearly observed in the galaxy group NGC~507 \citep{brienza2022}. 
Finally, it is fully possible that the source plasma is not passively ageing but that it has instead been re-energised (or that the thermal component evident in X-rays is confining the radio-emitting regions) and so, the timescales derived through the radiative ageing analysis are a lower limit to the actual plasma age. 

The radio emission observed in A1318 is very much reminiscent of the morphology seen in 3C~338 \citep{Burns1983} associated with the BCG of Abell~2199, where a bright ridge is detected connecting two old radio lobes but displaced to the south with respect to the nucleus of the parent galaxy. \cite{Nulsen2013} also studied that system and proposed that the ridge and the old lobes were produced by a previous AGN outburst associated with the BCG and later got  displaced as a consequence of the ICM sloshing motion initiated by a minor merger event.
The separation between the ridge and the currently active AGN in A2129 is much smaller ($\sim$6 kpc) with respect to what observed in A1318 ($\sim$170 kpc). However, this scenario might also hold for A1318 and should be taken into account in our interpretation of the source kinematics and morphology. We recall that, for example, in NGC~507 \citep{brienza2022} the old remnant plasma, which is likely to be transported by sloshing as well, was found at a distance of $\sim$80 kpc from the BGG. 

Unfortunately, the available X-ray image is not deep enough to allow a detailed analysis of the IGrM morphology. However, a hint to the fact that the IGrM is not uniformly distributed around the remnant source in A1318 comes from the morphology of the two lobes. In particular, components A and B have more sharply delineated boundaries than component C, which appears more diffuse, suggesting confinement in the former (component A seems to be in a cavity defined X-ray emission around it). Most importantly, the filaments observed in component B are not at all typical of a passively evolving remnant plasma; instead, they are more similar to what has been observed in Phoenix sources, again consistent with the idea of a dynamically disturbed system (e.g. \citealp{Mandal2019}). 
They have curved radio spectra and overall flatter spectral indices with respect to the surrounding plasma (Figure \ref{fig:A1318_spix-LBA-HBA-GMRT-WSRT33_47asec}), suggesting that particles are being compressed or re-energised there. 

Similar filamentary ridges possibly related to AGN remnant plasma are observed in other systems as well, for example in Nest~200047 \citep{Brienza2021}, in 3C~442 \citep{Hardcastle2007}, in the 'kite'-like source reported on by \cite{Ignesti2020}, in NGC~507 \citep{brienza2022}, and are all considered to arise from the interaction between the AGN radio plasma and the surrounding thermal medium. Simulations also support the creation of such structures as a consequence of the evolution of the old AGN plasma, combined with minor dynamical interactions of the system (e.g. \citep{Vazza2021, Vazza2023}.

If the filaments are being observed (in projection) to be part of component B (while in reality they are standalone structures) or if component B itself can be considered a large filament (note its sharp brightness cutoff at the north and south edge) with substructures, then they are very much resembling lobe-connected filaments in an active AGN, reported by \cite{Rudnick2022} and \cite{Knowles2022}, for instance, where components A and C are the (remnant) lobes it connects. The study of \cite{Knowles2022} may indicate that companion galaxies to the AGN host galaxy (present in our case as well) may induce filament formation by influencing the host dynamics and jet collimation. The caveat here is that the dynamical and emission timescales of A1318 are substantially different. 

In conclusion, we report the discovery of a spatially extended AGN remnant associated with the galaxy group Abell~1318. We performed an radio morphological, spectral, and ageing analysis and we conclude that we are seeing a multi-epoch AGN activity spanning (at least) 120 Myr. The radio morphology is indicative of filamentary structure reported recently in the literature as a distinct and yet unexplained phenomenon. This may possibly arise as the consequence of a complex 'group-weather' in this system.

New deep X-ray observations are now crucial to investigate the possible interaction between the X-ray emitting baryons and their interaction with the observed non-thermal radio plasma, as well as more thorough optical spectroscopy to characterise the surrounding galaxies. In addition, deep high-resolution radio images at frequencies of >~ 1.4 GHz will also be crucial to probe the radio spectrum of the source over a broader frequency range, as well as to investigate the polarisation properties of the observed filaments and understand their origin.

\begin{acknowledgements}

This paper is based (in part) on data obtained with the International LOFAR Telescope (ILT) under project codes LC3\_008 and LT16\_004. LOFAR \citep{RefWorks:157} is the Low Frequency Array designed and constructed by ASTRON. It has observing, data processing, and data storage facilities in several countries, that are owned by various parties (each with their own funding sources), and that are collectively operated by the ILT foundation under a joint scientific policy. The ILT resources have benefited from the following recent major funding sources: CNRS-INSU, Observatoire de Paris and Université d'Orléans, France; BMBF, MIWF-NRW, MPG, Germany; Science Foundation Ireland (SFI), Department of Business, Enterprise and Innovation (DBEI), Ireland; NWO, The Netherlands; The Science and Technology Facilities Council, UK; Ministry of Science and Higher Education, Poland; IstitutoNazionale di Astrofisica (INAF). This research has made use of the University of Hertfordshire high-performance computing facility (\href{https://uhhpc.herts.ac.uk/}{https://uhhpc.herts.ac.uk/}) and the LOFAR-UK compute facility, located at the University of Hertfordshire and supported by STFC [ST/P000096/1]\\
We thank the staff of the GMRT that made these observations possible. GMRT is run by the National Centre for Radio Astrophysics of the Tata Institute of Fundamental Research.\\
This work makes use of data from the Apertif system installed at the Westerbork Synthesis Radio Telescope owned by ASTRON. ASTRON, the Netherlands Institute for Radio Astronomy, is an institute of the Dutch Research Council (“De Nederlandse Organisatie voor Wetenschappelijk Onderzoek, NWO). \href{http://hdl.handle.net/21.12136/B014022C-978B-40F6-96C6-1A3B1F4A3DB0}{http://hdl.handle.net/21.12136/B014022C-978B-40F6-96C6-1A3B1F4A3DB0}\\
This research has made use of the NASA/IPAC Extragalactic Database (NED), which is operated by the Jet Propulsion Laboratory, California Institute of Technology, under contract with the National Aeronautics and Space Administration.\\
This research has made use of APLpy, an open-source plotting package for Python hosted at \href{http://aplpy.github.com}{http://aplpy.github.com}\\
This research has made use of "Aladin sky atlas" developed at CDS, Strasbourg
Observatory, France.\\
FdG acknowledges support from the Deutsche Forschungsgemeinschaft under Germany's Excellence Strategy - EXC 2121 "Quantum Universe" - 390833306. FdG acknowledges the support of Fondazione Cariplo and Fondazione CDP, grant \textnumero\ 2022-1801.\\
KR acknowledges financial support from the ERC Starting Grant "MAGCOW" \textnumero\ 714196.\\
MB acknowledges support from the Deutsche Forschungsgemeinschaft under Germany's Excellence Strategy - EXC 2121 "Quantum Universe" - 390833306.\\
BA acknowledges funding from the German Science Foundation DFG, within the Collaborative Research Center SFB1491 "Cosmic Interacting Matters - From Source to Signal".\\
M.Brienza acknowledges financial support from the agreement ASI-INAF n. 2017-14-H.O and from the PRIN MIUR 2017PH3WAT "Blackout".\\
DV acknowledges support from the Netherlands eScience Center (NLeSC) under grant ASDI.15.406\\
LCO acknowledges funding from the European Research Council under the European Union's Seventh Framework Programme (FP/2007-2013)/ERC Grant Agreement \textnumero\ 617199.\\
FM acknowledges support by the National Aeronautics and Space Administration (NASA) grants GO9-20083X, GO0-21110X and GO1-22087X.\\
KMH acknowledges financial support from the grant CEX2021-001131-S funded by MCIN/AEI/ 10.13039/501100011033 from the coordination of the participation in SKA-SPAIN funded by the Ministry of Science and Innovation (MCIN); from grant PID2021-123930OB-C21 funded by MCIN/AEI/ 10.13039/501100011033 by “ERDF A way of making Europe” and by the "European Union"; and funding from the ERC under the European Union’s Seventh Framework Programme (FP/2007–2013)/ERC Grant Agreement \textnumero\ 291531 (‘HIStoryNU’).\\
SAOImageDS9 development has been made possible by funding from the {\it Chandra} X-ray Science Center (CXC), the High Energy Astrophysics Science Archive Center (HEASARC) and the JWST Mission office at Space Telescope Science Institute \citep{Joye2003}.\\
This research has made use of data obtained from the high-energy Astrophysics Science Archive Research Center (HEASARC) provided by NASA’s Goddard Space Flight Center. We acknowledge the use of NASA's SkyView facility (\href{http://skyview.gsfc.nasa.gov}{http://skyview.gsfc.nasa.gov}) located at the NASA Goddard Space Flight Center.\\
This research has made use of datasets or services made available by the Infrared Science Archive (IRSA) at IPAC, which is operated by the California Institute of Technology under contract with the National Aeronautics and Space Administration.\\
TOPCAT and STILTS astronomical software \citep[\href{http://www.starlink.ac.uk/topcat/}{http://www.starlink.ac.uk/topcat/ ;}][]{Taylor2005} were used for the preparation and manipulation of some of the tabular data and images.

\end{acknowledgements}

\bibliographystyle{A1318}
\bibliography{A1318}

\begin{appendix}

\section{LoTSS DR2 catalog sources}
\label{app:int}
\begin{minipage}{\textwidth}
\centering
\noindent \captionof{table}{\small LoTSS DR2 catalog sources and their associated optical identification (where applicable) for the aperture marked with a green circle shown in Figure \ref{fig:A1318_composite}.}
\label{table:A1318_field_sources}
\small
\begin{tabular}{l l c c c c c }
\hline\hline\\
\small LoTSS ID & \small Optical ID & \small $ m_{\mathrm{R}} $ & RA [hh:mm:ss.ss] & DEC [dd:mm:ss.ss] & \small S$^{144}_{int}$ [mJy] & \small z \\
\hline\\
ILTJ113525.17+550438.5 & 1737p550\_0003133 & 17.17 & 11:35:25.16 & 55:04:38.55 & $ 2.27 \pm 0.28 $ & 0.10508 \\
ILTJ113527.19+550816.7 & 1736p552\_0000233 & 21.65 & 11:35:27.19 & 55:08:16.72 & $ 0.58 \pm 0.26 $ & \\
ILTJ113527.26+550502.2 & & & 11:35:27.26 & 55:05:02.23 & $ 2.57 \pm 0.33 $ & \\
ILTJ113527.50+550538.1 & 1737p550\_0003433 & 18.66 & 11:35:27.50 & 55:05:38.11 & $ 3.11 \pm 0.49 $ & 0.11745 \\
ILTJ113529.20+550457.4 & 1737p550\_0003265 & 20.62 & 11:35:29.20 & 55:04:57.47 & $ 1.50 \pm 0.34 $ & 0.48930 \\
ILTJ113541.08+550734.5 & 1741p552\_0000050 & 20.47 & 11:35:41.08 & 55:07:34.56 & $ 0.57 \pm 0.50 $ & \\
ILTJ113541.27+550334.4 $^{\tablefootmark{a}}$ & 1737p550\_0002937 & 21.74 & 11:35:41.27 & 55:03:34.42 & $ 13.00 \pm 0.55 $ & 0.65936 \\
ILTJ113541.75+550925.5 & 1741p552\_0000501 & 16.66 & 11:35:41.75 & 55:09:25.51 & $ 1.09 \pm 0.11 $ & 0.05808 \\
ILTJ113542.91+550500.4 & & & 11:35:42.91 & 55:05:00.45 & $ 1.18 \pm 0.75 $ & \\
ILTJ113545.14+550924.7 & & & 11:35:45.14 & 55:09:24.78 & $ 0.88 \pm 0.22 $ & \\
ILTJ113548.47+550714.4 & & & 11:35:48.47 & 55:07:14.40 & $ 0.59 \pm 0.33 $ & \\
ILTJ113548.98+550123.9 & & & 11:35:48.98 & 55:01:23.95 & $ 0.80 \pm 0.13 $ & \\
ILTJ113551.02+550357.4 & 1741p550\_0002869 & 16.43 & 11:35:51.02 & 55:03:57.49 & $ 0.44 \pm 0.12 $ & 0.05843 \\
ILTJ113552.68+550124.6 & 1741p550\_0002308 & 20.58 & 11:35:52.68 & 55:01:24.62 & $ 0.34 \pm 0.14 $ & \\
ILTJ113554.02+550849.9 & 1741p552\_0000387 & 24.61 & 11:35:54.02 & 55:08:49.95 & $ 2.98 \pm 0.25 $ & 1.13452 \\
ILTJ113554.26+550927.3 & 1741p552\_0000516 & 19.17 & 11:35:54.26 & 55:09:27.31 & $ 0.81 \pm 0.17 $ & 0.42365 \\
ILTJ113554.64+550239.5 & & & 11:35:54.63 & 55:02:39.52 & $ 1.46 \pm 0.11 $ & \\
ILTJ113556.61+550211.7 & 1741p550\_0002407 & 22.39 & 11:35:56.61 & 55:02:11.77 & $ 0.43 \pm 0.13 $ & \\
ILTJ113558.29+550200.2 & 1741p550\_0002406 & 17.19 & 11:35:58.29 & 55:02:00.28 & $ 1.51 \pm  0.16 $ & 0.09554 \\
ILTJ113559.40+550621.7 $^{\tablefootmark{b}}$ & 1741p550\_0002727 & 14.11 & 11:35:59.40 & 55:06:21.76 & $ 623.55 \pm 13.89 $ & 0.05707 \\
ILTJ113601.88+550246.7 $^{\tablefootmark{c}}$ & 1741p550\_0002598 & 19.74 & 11:36:01.88 & 55:02:46.79 & $ 1.23 \pm 0.18 $ & 0.69200 \\
ILTJ113602.60+550630.1 & 1741p550\_0003469 & 19.40 & 11:36:02.60 & 55:06:30.13 & $ 1.89 \pm  0.83 $ & \\
ILTJ113603.86+550350.2$^{\tablefootmark{f}}$ & 1741p550\_0002726 & 14.75 & 11:36:03.86 & 55:03:50.20 & $ 0.57 \pm 0.13 $ & 0.05463 \\
ILTJ113605.28+550856.1 & 1741p552\_0000406 & 21.57 & 11:36:05.28 & 55:08:56.15 & $ 2.12 \pm 0.31 $ & 0.61926 \\
ILTJ113614.68+550443.0 & 1741p550\_0003093 & 23.06 & 11:36:14.68 & 55:04:43.08 & $ 0.51 \pm 0.15 $ & 1.12066 \\
ILTJ113616.15+550322.0 & & & 11:36:16.15 & 55:03:22.02 & $ 4.26 \pm 0.39 $ & \\
ILTJ113618.55+550447.4 & 1741p550\_0003088 & 19.84 & 11:36:18.55 & 55:04:47.44 & $ 0.84 \pm 0.25 $ & 0.32570 \\
ILTJ113618.57+550316.6 & 1741p550\_0002767 & 20.90 & 11:36:18.57 & 55:03:16.69 & $ 0.34 \pm 0.10 $ & \\
ILTJ113618.96+550700.1 $^{\tablefootmark{d}}$ & 1741p550\_0003598 & 20.36 & 11:36:18.96 & 55:07:00.13 & $ 9.46 \pm 0.36 $ & 0.50105 \\
ILTJ113620.24+550701.6 & & & 11:36:20.24 & 55:07:01.60 & $ 0.91 \pm 0.44 $ & \\
ILTJ113622.26+550453.4 & & & 11:36:22.26 & 55:04:53.47 & $ 0.65 \pm 0.11 $ & \\
\hline
\end{tabular}
\tablefoot{
Optical identifications are made using the DESI Legacy Imaging Surveys. (1) LoTSS catalog ID; (2) Legacy ID; (3) r band magnitude of the optical association; (4) and (5) right ascension and declination of the radio source (mean position, J2000); (6) Flux density at 144 MHz; and (7) Photometric or spectroscopic (SDSS, if available) redshift.
\tablefoottext{a}{Radio source marked E in Figure \ref{fig:A1318_composite}.}
\tablefoottext{b}{Target source (components A, B, C, D, and F) associated with the BGG.}
\tablefoottext{c}{QSO, X-ray source 2RXS\,J113600.3+550210}
\tablefoottext{d}{QSO}
\tablefoottext{f}{Elliptical galaxy companion of BGG}
}

\end{minipage}

\clearpage

\begin{minipage}{\textwidth}
\section{Spectral index and spectral age error maps}
\label{app:err}

Below, we present the spectral index error maps associated with the corresponding figures in the main text.
    \centering
    \includegraphics[width=0.5\textwidth]{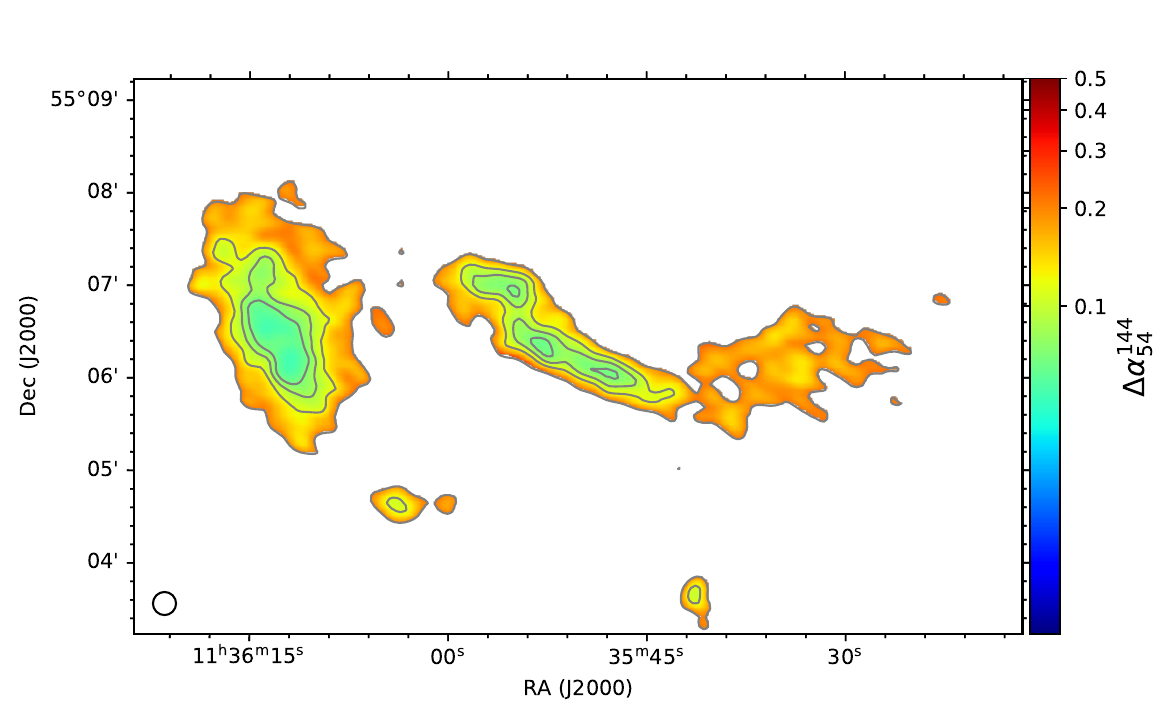}%
    \includegraphics[width=0.5\textwidth]{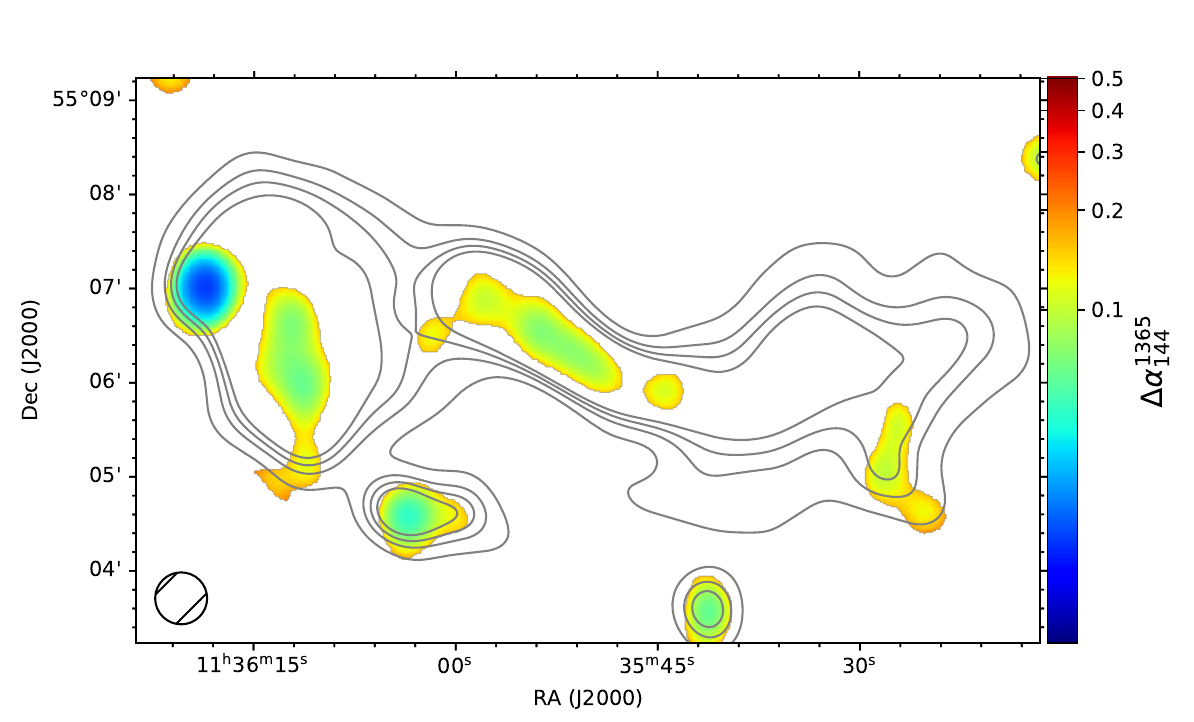}\\
    \includegraphics[width=0.5\textwidth]{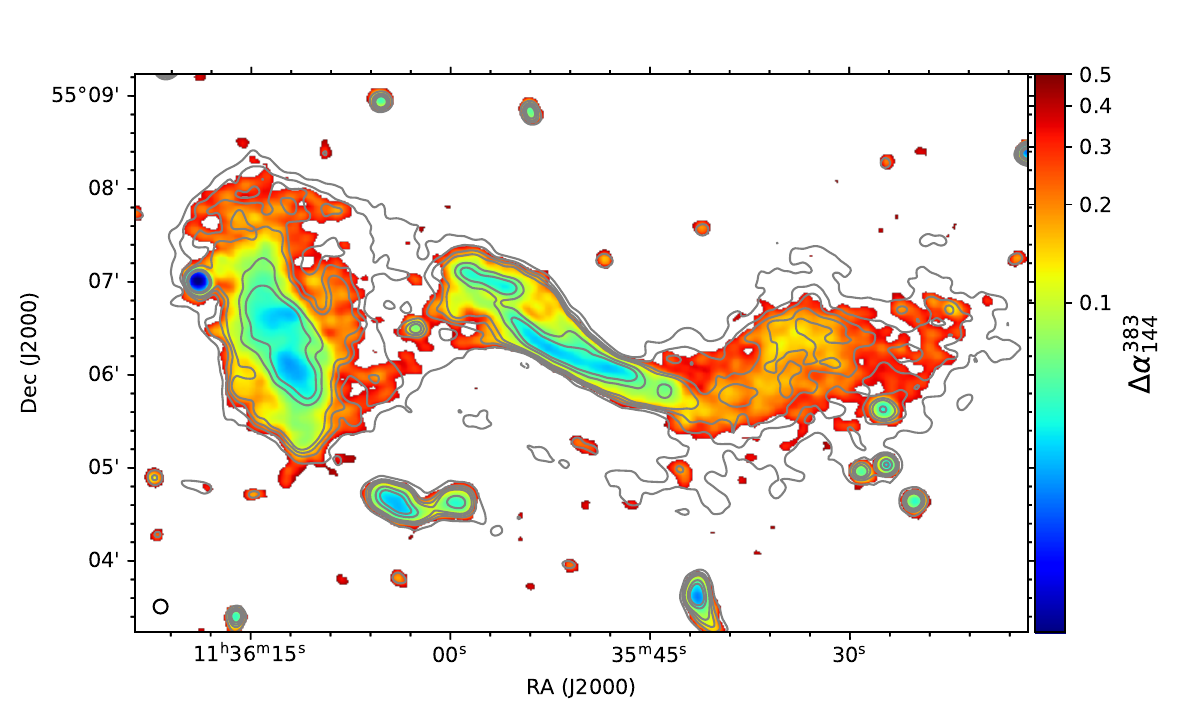}%
    \includegraphics[width=0.5\textwidth]{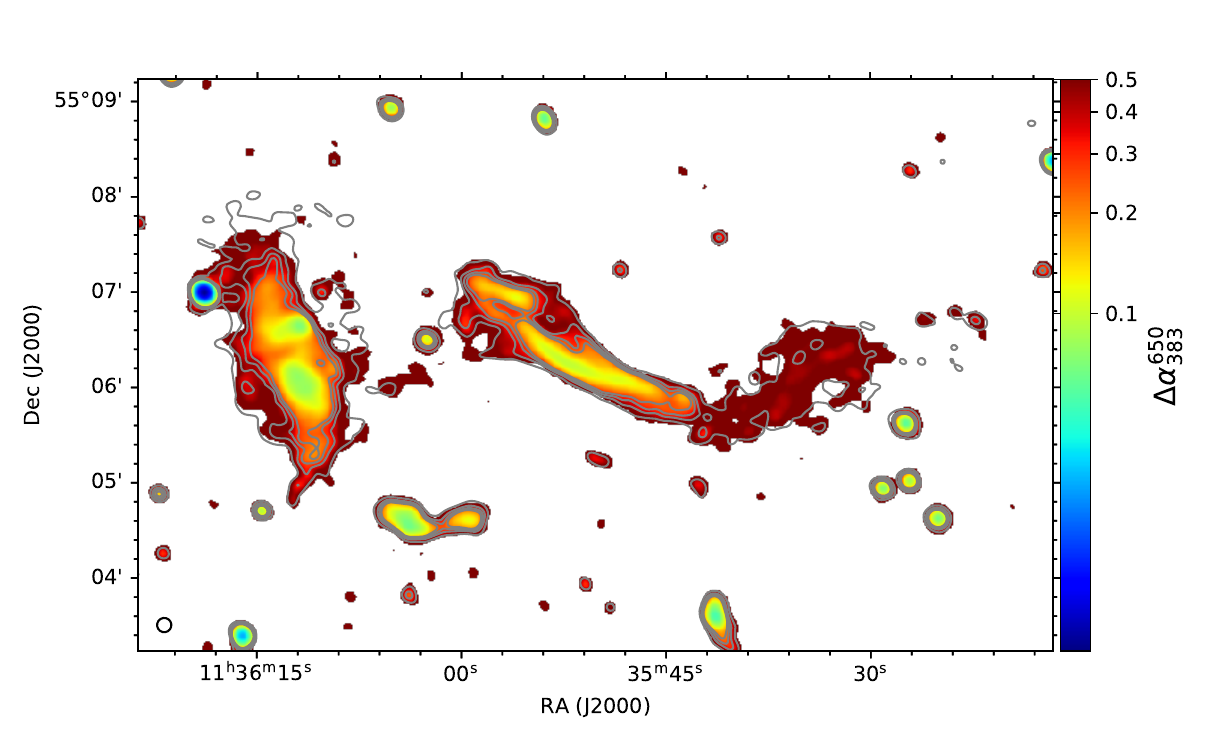}
    \captionof{figure}{$\alpha_{54}^{144}$ (LOFAR/LBA - LOFAR/HBA) spectral index error (top left panel) and $\alpha_{54}^{1365}$ (LOFAR/HBA - APERTIF) spectral index limits error map (top right panel). $\alpha_{144}^{383}$ (bottom left) and $\alpha_{383}^{650}$ (bottom right) LOFAR/HBA/GMRT spectral index error maps. Map parameters identical to those reported in Figure \ref{fig:A1318_spix-LBA-HBA-GMRT-WSRT33_47asec}.}
    \label{fig:A1318_spixerr-LBA-HBA-GMRT-WSRT33_47asec}
\end{minipage}

\begin{minipage}{\textwidth}
    \centering
    \includegraphics[width=0.5\textwidth]{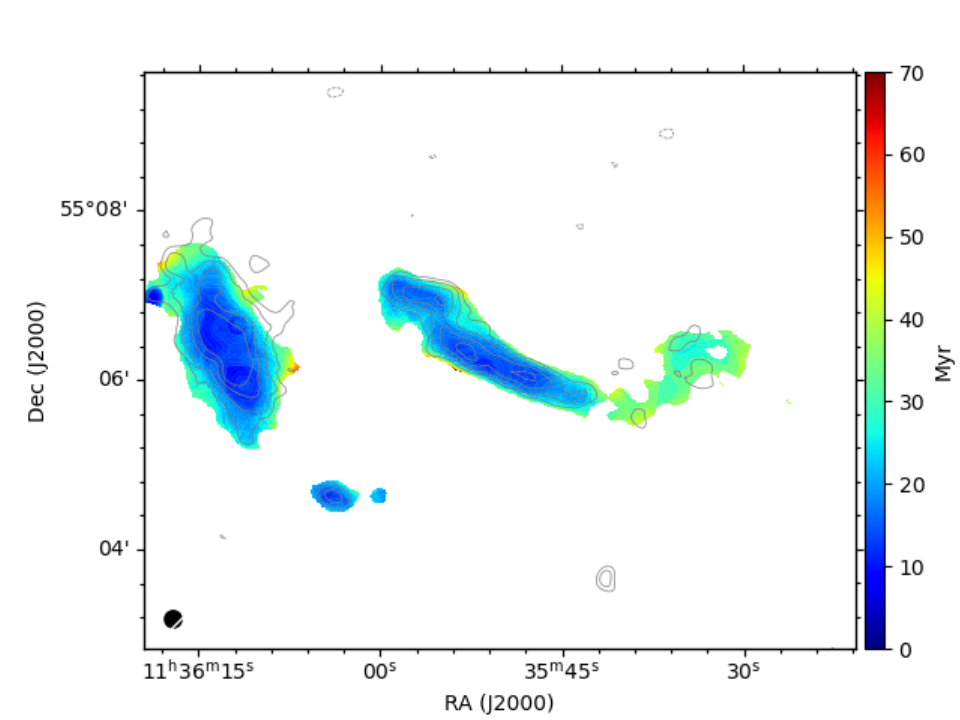}%
    \includegraphics[width=0.5\textwidth]{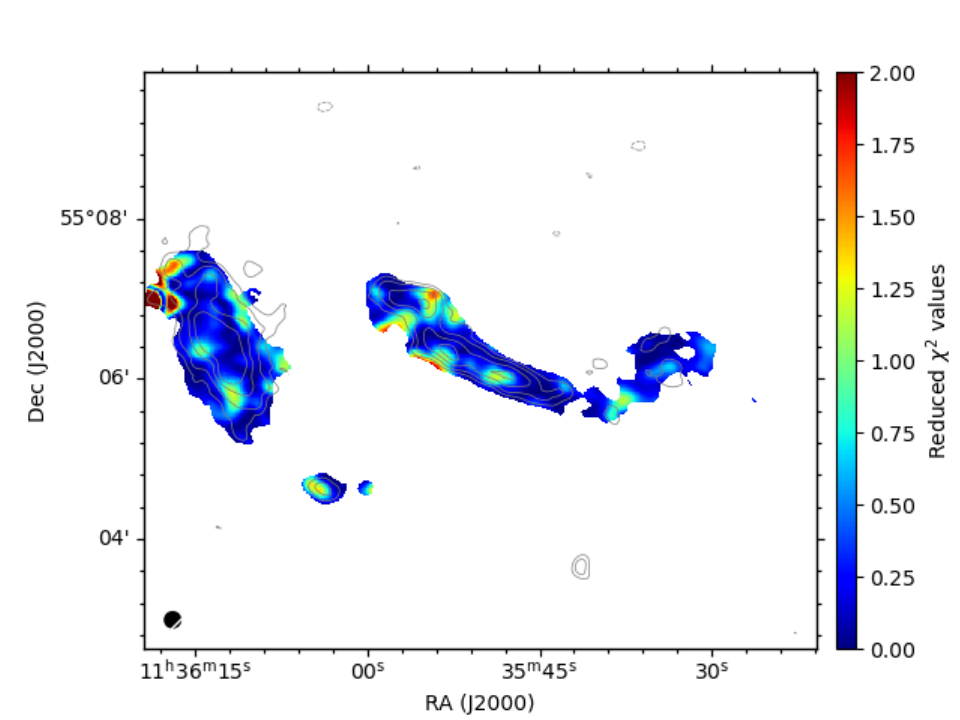}    \captionof{figure}{Spectral age error (left) and goodness of model fit (right) maps.}
\end{minipage}

\end{appendix}

\end{document}